\documentclass[12pt]{article}

\usepackage{graphicx}
\usepackage{makecell}
\usepackage{booktabs}
\usepackage{array}
\usepackage{fullpage}
\usepackage{url}
\usepackage{algorithm}
\usepackage{algorithmic}
\usepackage{bm}
\usepackage{smile}
\usepackage{wrapfig}
\usepackage{lipsum}
\usepackage{mathrsfs}
\usepackage{dsfont}
\usepackage{titling}
\usepackage{epstopdf}
\usepackage{multirow}
\usepackage{float}
\usepackage{rotating}

\usepackage[usenames,dvipsnames,svgnames,table]{xcolor}
\usepackage{hyperref}
\hypersetup{colorlinks=true,
            linkcolor=blue,
            urlcolor=blue,
            citecolor=blue}

\providecommand{\norm}[1]{\|#1\|}

\makeatletter

\newcommand{\Rmnum}[1]{\uppercase\expandafter{\romannumeral #1}}
\makeatother

\usepackage{xr}
\usepackage{geometry}
\usepackage{setspace}
\begin{document}

\title{Optimal and Safe Estimation for High-Dimensional Semi-Supervised Learning}
\author{
Siyi Deng\thanks{Department of Statistics and Data Science, Cornell University, Ithaca, NY 14850, USA; e-mail: \texttt{sd847@cornell.edu}.}~~~~~~
Yang Ning\thanks{Department of Statistics and Data Science, Cornell University, Ithaca, NY 14850, USA; e-mail: \texttt{yn265@cornell.edu}.}~~~~~~
Jiwei Zhao\thanks{Department of Biostatistics and Medical Informatics, University of Wisconsin-Madison, Madison, WI 53726, USA; e-mail: \texttt{jiwei.zhao@wisc.edu}.}~~~~~~
Heping Zhang\thanks{Department of Biostatistics, Yale University, New Haven, CT 06511, USA; e-mail: \texttt{heping.zhang@yale.edu}.}
}

\date{\today}

\maketitle
\begin{abstract}
We consider the estimation problem in high-dimensional semi-supervised learning.
Our goal is to investigate when and how the unlabeled data can be exploited to improve the estimation of the regression parameters of linear model in light of the fact that such linear models may be misspecified in data analysis.
We first establish the minimax lower bound for parameter  estimation  in the semi-supervised setting, and show that this lower bound cannot be achieved by supervised estimators using the labeled data only.
{
We propose an optimal semi-supervised estimator that can attain this lower bound and therefore improves the supervised estimators, provided that the conditional mean function can be consistently estimated with a proper rate.
}
We further propose a safe semi-supervised estimator.
We view it safe, because this estimator is always at least as good as the supervised estimators.
We also extend our idea to the aggregation of multiple semi-supervised estimators caused by different misspecifications of the conditional mean function.
Extensive numerical simulations and a real data analysis are conducted to illustrate our theoretical results.
\end{abstract}

{\bf Key Words:}
High dimensionality;
model aggregation;
model misspecification;
optimal estimation;
safe estimation;
semi-supervised learning.

\newpage

\setcounter{equation}{0}

\section{Introduction}\label{sec_intro}

Semi-supervised learning is an emerging research area in statistics and machine learning \citep{zhu2005semi, chapelle2006semi}, and can have a great potential in electronic health records (EHR) based studies for clinical research.
In these types of studies, one major challenge is the lack of gold-standard health outcomes or phenotypes \citep{kohane2011using}.
The validated phenotypes are often obtained by manual chart reviews that are prohibitively expensive \citep{liao2010electronic}; therefore,
only can a very small subset of patients be annotated by  experts in reality. For the rest
of the
patients, their covariate information, often high-dimensional \citep{weisenthal2018predicting, gensheimer2019automated, abdullah2020visual}, is only available.

Developing efficient statistical methods to analyze such data is a timely and important problem.
Let $Y$ denote the outcome variable and $X$ the $p$-dimensional covariates. In addition to $n$ independent and identically distributed (i.i.d.) samples $(Y_1, X_1),...(Y_n, X_n)\sim (Y, X)$, we also observe $N$ i.i.d. data consisting of only covariates, $X_{n+1},...,X_{N+n}\sim X$. Following the convention, the former is referred  to as labeled data and the latter is called unlabeled data. We also denote $\bY=(Y_1,...,Y_n)^T\in\RR^n$ and $\bX=(X_1,...,X_n)^T \in\RR^{n\times p}$ the outcomes and covariates from the labeled  data, and $\Tilde{\bX}=(X_1,...,X_{N+n})^T \in\RR^{(N+n)\times p}$ the covariates from both. In this work, we focus on high-dimensional regression problems; namely, $p$ can be much larger than $n$. The size of the unlabeled data $N$ is allowed but not required to be larger than $n$.

We consider the so-called assumption lean regression framework \citep{Buja,Assumption},
\begin{equation} \label{eq_model}
Y=f(X)+\epsilon,
\end{equation}
where $f(X)=E(Y|X)$ is the unknown conditional mean function, $\epsilon$ is the random error independent of $X\in \RR^p$ with $\EE(\epsilon)=0$, $\EE(\epsilon^2)=\sigma^2$, {and $\sigma^2$ is an unknown parameter}. We consider the random design and assume that $X$ and $Y$ are centered with $\EE(X)=0$ and $\EE(f(X))=0$. On one hand, we would like to put as  fewer assumptions  as possible on $f(X)$ to enable model flexibility. On the other hand, for the purpose of interpretability, we often fit simple parametric models such as linear regression to explain the association between $Y$ and $X$. To  meet both ends, we consider linear regression as a working model where the true data generating process follows (\ref{eq_model}). Since $\EE[(Y-X^T\btheta)^2]=\EE[(f(X)-X^T\btheta)^2]+\sigma^2$, the regression coefficients in a linear model correspond to the $L_2(\PP)$ projection of $f(X)$ onto the linear space spanned by $X$, i.e.,
$$
\btheta^*=\arg\min_{\theta\in \mathbb{R}^p} \EE[(f(X)-X^T\btheta)^2]
$$
that delineates the linear dependence between $Y$ and $X$.
We do not include intercept in $\btheta^*$ simply because $\EE(X)=0$ and $\EE(f(X))=0$.
Our goal here is to estimate the high dimensional parameter $\btheta^*$.

In the supervised setting with  $n$ labeled observations, a number of penalized estimators have been proposed to estimate $\btheta^*$, such as lasso \citep{lasso} and Dantzig selector \citep{candes2007}.
While significant progress has been made towards understanding the estimation in the fully supervised setting, there is very limited research in the semi-supervised setting. It is important to observe that under (\ref{eq_model}), since the linear regression is the working model, the covariate $X$ is no longer the ancillary statistic for the regression parameter $\btheta^*$. Therefore,
the covariate $X$ in the unlabeled data, usually with a much bigger sample size than the labeled data, is informative and may be beneficial for estimating $\btheta^*$.

Our first contribution is to establish the minimax lower bound for estimating $\btheta^*$ in the semi-supervised setting.
{
In particular, to derive this lower bound, we assume $f(\cdot)$ is unknown but belongs to some specific model class, such that methods for estimating $f(\cdot)$ are available in the existing literature.
}
Based on the lower bound, when $N$ is sufficiently large, the error term due to the model misspecification or equivalently the nonlinearity of $f(\cdot)$ becomes negligible. This reveals one potential benefit from using the unlabeled data in that the estimation of $\btheta^*$ can be more robust to the model misspecification. Moreover, we show that the fully supervised estimators (e.g., lasso and Dantzig selector) do not attain this lower bound.
Our second contribution is to propose a new semi-supervised estimator that achieves this lower bound under some conditions. In Theorem \ref{mainthm}, we show that the rate of our estimator depends on whether the unknown $f(\cdot)$ can be consistently estimated.
{
When $f(\cdot)$ belongs to some specific model class so that it can be consistently estimated with a proper rate,
}
the proposed estimator attains the  minimax lower bound up to some logarithmic factor, and therefore improves the rate of the supervised estimators. When $f(\cdot)$ is misspecified, however, the rate of our semi-supervised estimator becomes sub-optimal and may be even slower than the supervised estimators.
Our third contribution is to develop a general two-step refitting procedure that further improves the above semi-supervised estimator.
The resulting estimator is guaranteed to be no worse than the supervised estimators no matter $f(\cdot)$ is misspecified or not, and
remains minimax rate-optimal when $f(\cdot)$ belongs to some specific model class so that can be consistently estimated with a proper rate, hence it provides a safe use of the unlabeled data. Therefore, we call it the safe semi-supervised estimator.
In empirical studies one may encounter the situation that, while all misspecified, different estimates of $f(\cdot)$ are available.
We further extend the idea of creating safe semi-supervised estimator to the aggregation of multiple semi-supervised estimators under such a situation. The convergence rate of the aggregated estimator is guaranteed to be no worse than any of the un-aggregated semi-supervised estimators.
Overall, our goal is to exploit both safe and efficient use of the unlabeled data by developing semi-supervised estimators of $\btheta^*$ such that their convergence rates are faster, or at least no slower, than the standard supervised estimators (e.g., lasso and Dantzig selector).

\subsection{Related work}

In computer science, a large number of classification algorithms have been developed under semi-supervised setting, which mainly focus on data with discrete labels; see \cite{zhu2005semi,chapelle2009semi} for some surveys. Common assumptions such as manifold assumption and cluster assumption were made in the literature in order to obtain fast rate of convergence in classification \citep{cluster}. In non-parametric regression problem, \cite{wasserman} showed that unlabeled data do not always help to improve the  rate of the mean squared error, but with semi-supervised smoothness assumption the estimator with faster rate could be developed.

More recently, \cite{van2020survey} and \cite{yang2021survey} provided comprehensive surveys on many up-to-date developments in semi-supervised learning, especially with deep neural networks. For example, the methods named as Mixmatch \citep{mixmatch} and Fixmatch \citep{fixmatch} leverage unlabeled data through pseudo-labeling and consistency regularization to assist the prediction model. The self-training method, which trains a model to fit pseudo-labels predicted by previously learned models, has also been developed for semi-supervised learning \citep{xie2020self,chen2020self,wei2020theoretical}. The performance of many modern semi-supervised learning algorithms in some image classification tasks is compared and discussed by \cite{realistic}.

There are also progress considering how to make use of the unlabeled data to obtain an estimator with a smaller asymptotic variance, when the dimension $p$ is fixed and small. For example, \cite{zhang2019, azriel} and \cite{chakrabortty2018} investigated how to  incorporate the unlabeled data to improve the estimation efficiency for the population mean $\EE(Y)$ and regression coefficients in a working linear regression.

With high-dimensional data, \cite{earliest} proposed a transductive version of lasso and Dantzig selector in the semi-supervised setting. They showed that the transductive estimators have the same rate as the supervised ones. More recently, \cite{predictionloss} proposed a modified lasso estimator and showed that the excess risk of their estimator in prediction has the same rate of convergence as the supervised lasso estimator. These existing theoretical results neither confirm nor deny the existence of estimators with improved statistical rate when unlabeled data are available.
To the best of our knowledge, it remains an open problem of when and how one can develop a semi-supervised estimator with improved statistical rate by exploiting the available unlabeled data.
We bridge this gap by showing the minimax lower bound and proposing new semi-supervised estimators.

\subsection{Organization of the paper}

The rest of this paper is organized as follows. In Section~\ref{sec_lower}, we give the minimax lower bound for semi-supervised estimation. In Section~\ref{sec_upper}, we introduce the optimal semi-supervised estimator and its corresponding upper bound. In Section~\ref{sec_safe} we devote ourselves into the safe semi-supervised estimator, where we propose a two-step procedure regardless of the quality of the estimated conditional mean model. In Section~\ref{sec_safe2} we extend the idea to aggregation of multiple semi-supervised estimators caused by different
misspecifications of the conditional mean model.
Numerical experiments and a real data application are in Sections~\ref{sec_simu} and \ref{sec_data}, respectively.
All the technical proofs are contained in the Supplement.

\subsection{Notations}

Let $P_{X,Y}$ and $P_X$ denote the joint distribution of $(X,Y)$ and the marginal distribution of $X$, respectively.
For $v=(v_1,...,v_p)^{T} \in \mathbb{R}^p$,
we define $\norm{v}_0=|\textrm{supp}(v)|$ where $\textrm{supp}(v)=\{i: v_i\neq 0\}$ and $|A|$ is the cardinality of a set $A$,
$\norm{v}_q=(\sum_{i=1}^p |v_i|^q)^{1/q}$ for $1 \leq q < \infty$, and $\norm{v}_{\infty}=\max_{1\leq i \leq p} |v_i|$. Denote $v^{\otimes 2}=v v^T$.
For a matrix $\Mb=[M_{ij}]$, $\Mb_{i\cdot}$ and $\Mb_{\cdot j}$ denote the $i$-th row and $j$-th column respectively. Define  $\norm{\Mb}_{\max}=\max_{ij}|M_{ij}|$, $\norm{\Mb}_1=\max_{j}\sum_{i}|M_{ij}|$, $\norm{\Mb}_{\infty}=\max_{i}\sum_{j}|M_{ij}|$. If the matrix $\Mb$ is symmetric, then $\Lambda_{\min}(\Mb)$ and $\Lambda_{\max}(\Mb)$ are the minimal and maximal eigenvalues of $\Mb$. We denote $\Ib_p$ the $p\times p$ identity matrix. For $S\subseteq \{1,...,p\}$, let $v_S=\{v_k: k\in S\}$ and $ S^c$ be the complement of $S$.
{For matrix $\bX\in\RR^{n\times p}$ and index set $D\subseteq \{1,...,n\}$, $\bX_D=\{X_i:i\in D\}^T\in \RR^{|D|\times p}$.}
{
For a function $f$, let $\norm{f}_2=\sqrt{\EE[f(X)^2]}$ denote the $L_2(\PP)$ norm of $f$.
}

For two positive sequences $a_n$ and $b_n$, we write $a_n\asymp b_n$ if $C\leq a_n/b_n\leq C'$ for some $C,C'>0$. Similarly, we use $a\lesssim b$ to denote $a\leq Cb$ for some constant $C>0$. Given $a,b\in\RR$, let $a\vee b$ and $a\wedge b$ denote the maximum and minimum of $a$ and $b$.

\section{Minimax Lower Bound for Semi-Supervised Estimation}\label{sec_lower}

The semi-supervised learning setting refers to that we observe $n$ i.i.d. copies of $(Y, X)$ and additional $N$ i.i.d. copies of $X$, where the distributions of $X$ in both labeled and unlabeled data are the same.
{
In Theorem~\ref{mainthm} presented in Section~\ref{sec_upper}, we will rigorously show that  the unknown conditional mean function $f(\cdot)$ plays an important role in assessing the optimality of semi-supervised estimators.
However, the correct specification and consistent estimation of $f(\cdot)$ under high dimensionality is by no means a trivial problem.
In the literature, methods for consistently estimating $f(\cdot)$ with a proper rate are only available when $f(\cdot)$ belongs to some specific model class.
In that regard, when analyzing the minimax lower bound here, we assume that $f(\cdot)$ belongs to one of the following two model classes.
}

We first define the pairwise interaction model
$\cF_{\textrm{pairwise}}=\{\sum_{j=1}^p \gamma_j X_j+\sum_{1\leq j\leq k\leq p} \gamma_{jk} X_jX_k\}$,
where $\bgamma=(\gamma_1,...,\gamma_p,\gamma_{11},\gamma_{12},...,\gamma_{pp})\in\RR^{p+p(p+1)/2}$ are unknown parameters satisfying $\|\bgamma\|_0\leq \bar s$.
In practice, the conditional mean function $f(X)$ is usually nonlinear in $X$.
In $\cF_{\textrm{pairwise}}$, we account for the nonlinearity by incorporating the quadratic terms and the pairwise interactions.
To mitigate the model complexity, the parameter $\bgamma$ is assumed to be $\bar s$-sparse \citep{zhao2016analysis}.
We refer to Supplement \ref{app_sup} for further discussions.
Thus, we define the class of joint distributions of $(X,Y)$ as
$$\begin{aligned}
\mathcal{P}^{\textrm{pairwise}}_{\Phi,\sigma}=\{P_{X,Y}|~ Y = f(X)+\epsilon,~
&f(\cdot)\in \cF_{\textrm{pairwise}}, ~
\|\btheta^*\|_0\leq s, ~ \Var(\epsilon)=\sigma^2,\\
&\EE(f(X)-X^T\btheta^*)^2 \leq \Phi^2 , ~\textrm{and}~ P_X\in  \mathcal{P}_X \},
\end{aligned}$$
where $\btheta^*$ implicitly depends on the distribution $P_{X,Y}$, the parameter $s$ controls the sparsity of $\btheta^*$, $\mathcal{P}_X=\{P_X|~ \EE[X]=0, \Var(X_j)=1 ~\textrm{and}~ \Lambda_{\min}(\Cov(X))\geq C_{\min}>0$\} {with some constant $C_{\min}$}.
For notational simplicity, we write $\EE(\cdot)$ for $\EE_{P_{X,Y}}(\cdot)$. We note that, $\mathcal{P}^{\textrm{pairwise}}_{\Phi,\sigma}$ is indexed by two non-negative parameters $\Phi^2$ and $\sigma^2$, where the former controls the magnitude of model misspecification $f(X)-X^T\btheta^*$ or equivalently the nonlinearity of $f(X)$ in the second moment and the latter is the variance of $\epsilon$. In particular, we allow $\Phi^2$ to grow with $n$ in our framework.


The second model class we consider is the additive model $\cF_{\textrm{additive}}=\{\sum_{j=1}^p f_j(X_j)\}$, where $f_j$'s are unknown second-order-smooth functions \citep{lin2006,meier2009,huang2010,additive}. To ease the presentation,  we defer the definition of $\ell$-smooth functions to Supplement \ref{app_sup}. Similarly, we assume the number of nonzero functions is bounded by $\bar s$. Compared to $\cF_{\textrm{pairwise}}$, the additive model does not allow interactions among covariates but offers more flexibility in associating each component $X_j$ with $Y$.
Similar to $\mathcal{P}^{\textrm{pairwise}}_{\Phi,\sigma}$, we can define the class of distributions $\mathcal{P}^{\textrm{additive}}_{\Phi,\sigma}$, where we replace $f\in \cF_{\textrm{pairwise}}$ with $f\in \cF_{\textrm{additive}}$ in the  definition.

The following theorem offers the lower bound for the convergence rate of any estimator of $\btheta^*$ over the classes of distributions $\mathcal{P}^{\textrm{pairwise}}_{\Phi,\sigma}$ or $\mathcal{P}^{\textrm{additive}}_{\Phi,\sigma}$, in the semi-supervised setting.
Throughout the paper without causing confusion, we use $C, C', c_1,$ and  $c_2,$ etc. to denote generic constants whose values can change from time to time.

\begin{assumption} \label{assumption_low}
Assume that $s\log(p/s)\leq C n$ for some constant $C$, $4\leq s\leq (n-1)/4$, and the sparsity level in $\cF_{\textrm{pairwise}}$ and $\cF_{\textrm{additive}}$ satisfies $\bar s\geq s$.
\end{assumption}

\begin{theorem}\label{thm_lowerbound}
Under Assumption \ref{assumption_low}, we have that for any $1\leq q\leq \infty$,
\begin{equation} \label{eq_lowerbound}
\inf_{\hat\btheta} \sup_{P_{X,Y}\in \mathcal{P}^{\textrm{pairwise}}_{\Phi,\sigma}} \PP_{P_{X,Y}} \Big[\norm{\hat\btheta-\btheta^*}_q\geq  c_1s^{1/q}\Big(\Phi\sqrt{\frac{\log (p/s)}{n+N}}+\sigma\sqrt{\frac{\log (p/s)}{n}}\Big)\Big]>c_2,
\end{equation}
where $\inf_{\hat\btheta}$ denotes the infimum over all estimators based on the labeled data $(Y_i, X_i)$ for $1\le i\le n$ and unlabeled data $X_i$ for $n+1\le i\le n+N$, and $c_1$ and $c_2$ are some positive constants. Here we denote $s^{1/\infty}=1$. In addition, the same lower bound (\ref{eq_lowerbound}) holds when we replace $P_{X,Y}\in \mathcal{P}^{\textrm{pairwise}}_{\Phi,\sigma}$ with $P_{X,Y}\in \mathcal{P}^{\textrm{additive}}_{\Phi,\sigma}$.
\end{theorem}

\begin{remark}\label{rem_lower}
{
The lower bound (\ref{eq_lowerbound}) in Theorem~\ref{thm_lowerbound} is obtained by restricting $f(\cdot)$ to be in either $\cF_{\textrm{pairwise}}$ or $\cF_{\textrm{additive}}$.
Indeed, this is a stronger result than the case without such a restriction.
Theorem~\ref{thm_lowerbound} implies, if one considers
the class of joint distributions of $(X,Y)$ as
$$\begin{aligned}
\mathcal{P}_{\Phi,\sigma}=\{P_{X,Y}|~ Y = f(X)+\epsilon,~
&\norm{f}_2<\infty, ~
\|\btheta^*\|_0\leq s, ~ \Var(\epsilon)=\sigma^2,\\
&\EE(f(X)-X^T\btheta^*)^2 \leq \Phi^2 , ~\textrm{and}~ P_X\in  \mathcal{P}_X \}
\end{aligned}$$
without specifying the model class of $f(\cdot)$, the minimax lower bound remains the same as in (\ref{eq_lowerbound}).
}
This lower bound consists of two components. Up to some absolute constants, the first term $s^{1/q}\Phi\sqrt{{\log (p/s)}/(n+N)}$ corresponds to the error  due to potential model misspecification and the second term $s^{1/q}\sigma\sqrt{{\log (p/s)}/{n}}$ comes from the uncertainty inherited from the randomness of the error $\epsilon$, which always exists even if the regression function is linear $f(X)=X^T\btheta^*$. In this case, we have $\Phi=0$ and the lower bound agrees with the existing result for sparse linear regression \citep{verz,bellec2018slope}.
\end{remark}

\begin{remark}\label{rem_lower2}
The sample size of the unlabeled data $N$ plays an important role in the lower bound (\ref{eq_lowerbound}).
In Supplement \ref{app_phi}, we show that under some conditions, $\Phi^2\asymp s$ so $\Phi\to\infty$ as the sparsity grows and $\frac{\Phi}{\sigma}\sqrt{\frac{n}{n+N}}\rightarrow \infty$ may happen.
In this case, the dominating term in the lower bound $s^{1/q}\Phi\sqrt{{\log (p/s)}/(n+N)}$ can be reduced as $N$ increases. If $N$ is sufficiently large such that $\frac{\Phi}{\sigma}\sqrt{\frac{n}{n+N}}\rightarrow c<\infty$, the lower bound attains its minimum $s^{1/q}\sigma\sqrt{{\log (p/s)}/{n}}$, which can be viewed as the irreducible error in the semi-supervised setting since a further increase of $N$ would no longer decrease the lower bound.
As an illustration, we plot the lower bound in Figure \ref{fig_bound} of Supplement~\ref{sec:rateplot}.
\end{remark}

Before delving into our proposed estimators, we briefly summarize some known properties of supervised estimators which shall be useful later.
The supervised Dantzig selector is defined as
\begin{equation}
\label{dantini}
\hat\btheta_D=\arg\min \|\btheta\|_1,~~\textrm{s.t.}~~ \Big\|\frac{1}{n}\sum_{i=1}^n (Y_i-X_i^T\btheta)X_i\Big\|_\infty \leq \lambda_D,
\end{equation}
where $\lambda_D$ is a tuning parameter.  It is shown in Lemma \ref{bdant} of Supplement \ref{app_mainproof} that with high probability
\begin{equation}\label{eq_ratehat}
\|\hat\btheta_D-\btheta^*\|_1 = O_p\left\{s(\Phi+\sigma)\sqrt{\frac{\log p}{n}}\right\}.
\end{equation}
Under the condition $N\gg n$, the lower bound in (\ref{eq_lowerbound}) is strictly smaller in order than the upper bound (\ref{eq_ratehat}) if and only if $\Phi/\sigma\rightarrow\infty$.  In this case, the supervised estimator $\hat\btheta_D$ does not attain the lower bound and is thus sub-optimal in the minimax sense; see Figure \ref{fig_bound} in Supplement~\ref{sec:rateplot}.
Similarly, the supervised lasso estimator is defined as
\begin{equation}\label{eq_lasso}
\hat\btheta_L=\argmin_{\btheta \in \RR^p} \frac{1}{2n}\sum_{i=1}^n (Y_i-X_i^T\btheta)^2+\lambda_L\norm{\btheta}_1,
\end{equation}
where $\lambda_L$ is a tuning parameter.
The same upper bound as in (\ref{eq_ratehat}) can be derived similarly. The Dantzig selector and the lasso estimator are theoretically equivalent \citep{bickel2009}.
It turns out that,
our optimal semi-supervised estimator in Section~\ref{sec_upper} resembles the Dantzig selector (\ref{dantini}) while our safe semi-supervised estimator in Section~\ref{sec_safe} looks more similar to the lasso estimator (\ref{eq_lasso}).

\section{Optimal Semi-Supervised Estimator}\label{sec_upper}

\subsection{Motivation and the key step}\label{sec:motivation}
 To motivate our estimator, we first briefly explain how the convergence rate of $\hat\btheta_D$ in (\ref{dantini}) is derived. Following the standard argument in \cite{bickel2009}, the Dantizig selector satisfies $\|\hat\btheta_D-\btheta^*\|_1=O_p(s\lambda_D)$, where the tuning parameter $\lambda_D \gtrsim \|\frac{1}{n}\sum_{i=1}^n X_i(Y_i-X_i^T\btheta^*)\|_\infty$. In the proof of Lemma \ref{bdant}, we further show that $\|\frac{1}{n}\sum_{i=1}^n X_i(Y_i-X_i^T\btheta^*)\|_\infty\lesssim \sqrt{\frac{\log p}{n}}\{\EE(Y_i-X_i^T\btheta^*)^2\}^{1/2}$ with high probability. The desired bound (\ref{eq_ratehat}) is obtained by noting that
\begin{equation}\label{eq_var}
\EE(Y_i-X_i^T\btheta^*)^2=\EE(Y_i-f(X_i))^2+\EE(f(X_i)-X_i^T\btheta^*)^2\leq\sigma^2+\Phi^2.
\end{equation}
{
In view of (\ref{eq_ratehat}) and Remark \ref{rem_lower}, we see that the slow rate of $\hat\btheta_D$ is driven by the sup-norm of the score function $\|\frac{1}{n}\sum_{i=1}^n X_i(Y_i-X_i^T\btheta^*)\|_\infty$.
}

To find an estimator with the improved rate, our key idea is to construct a modified score function. To this end, we  decompose the score function of $\hat\btheta_D$ as
$$
\frac{1}{n}\sum_{i=1}^n X_i(Y_i-X_i^T\btheta^*)=\frac{1}{n}\sum_{i=1}^n X_i(Y_i-f(X_i))+\frac{1}{n}\sum_{i=1}^nX_i(f(X_i)-X_i^T\btheta^*).
$$
We propose to replace the last term as $\frac{1}{n+N}\sum_{i=1}^{n+N}X_i(f(X_i)-X_i^T\btheta^*)$, the sample average over  both labeled and unlabeled data. Apparently, it is a consistent estimator of $\EE[X_i(f(X_i)-X_i^T\btheta^*)]=-\frac{1}{2}\EE[\frac{\partial}{\partial\btheta}(f(X_i)-X_i^T\btheta^*)^2]$ with a faster rate. Thus, the unlabeled data can help estimate the expectation of the gradient of the model misspecification error. This explains why the unlabeled data may help in the case of misspecification. This leads to the following modified score function
\begin{equation}\label{eq_mscore}
\frac{1}{n}\sum_{i=1}^n X_i(Y_i-f(X_i))+\frac{1}{n+N}\sum_{i=1}^{n+N}X_i(f(X_i)-X_i^T\btheta^*)=\bar \bxi-\hat\bSigma_{n+N}\btheta^*,
\end{equation}
where $\hat\bSigma_{n+N}=\frac{1}{n+N}\sum_{i=1}^{n+N}X_i^{\otimes 2}$ and
\begin{equation}\label{eq_barxi}
\bar \bxi= \frac{1}{n}\sum_{i=1}^n X_iY_i-\frac{1}{n}\sum_{i=1}^n X_if(X_i)+\frac{1}{n+N}\sum_{i=1}^{n+N}X_if(X_i).
\end{equation}
Further insight on the modified score function can be found in Supplement \ref{app_sup}.

\subsection{Computation of $\bar \bxi$}\label{sec:computebxi}
To compute $\bar \bxi$, we need to find an estimator for $f(\cdot)$, the unknown conditional mean function.
In the rest of the paper, we use $\hat h(\cdot)$ to denote the estimate of the conditional mean function
and will discuss some examples in  Remark \ref{rem_consistency}. To account for the possible model misspecification of the unknown conditional mean function, we assume that there exists a function $h(\cdot)$ with $\|h\|_2<\infty$ such that the estimate $\hat h(\cdot)$ converges to $h(\cdot)$ in the $L_2(\PP)$ norm. We refer to $h$ as a conditional mean model. When the conditional mean model is correctly specified, we would expect that $f=h$ and $\hat h(\cdot)$ is consistent for $f(\cdot)$.

A serious challenge may arise from deriving the theoretical property of our proposed  semi-supervised estimator if we use all data to obtain $\hat h(\cdot)$ due to the dependence between the estimator $\hat h(\cdot)$ and the data $(X_i,Y_i)$ in the sample average from $\bar \bxi$. To bypass this challenge, we adopt the cross-fitting technique that was devised for semiparametric estimation problems \citep{bickel1982adaptive,schick1986asymptotically} as well as for high-dimensional data \citep{robins2017minimax,victor}. For notational simplicity, we denote by $D^*$ the labeled data and $D$ the full dataset. Without loss of generality, we split the labeled data $D^*$ into two folds $D^*_1$ and $D_2^*$ with size $n_1=n_2=n/2$. Similarly, we split the unlabeled data into two folds $U_1$ and $U_2$ with size $N_1=N_2=N/2$. Merging $U_1$ and $U_2$
with $D^*_1$ and $D_2^*$ respectively, we obtain two independent data sets $D_1=D_1^*\cup U_1$ and $D_2=D_2^*\cup U_2$. Next, for $j=\{1,2\}$, we train the estimator $\hat h^{-j}$ using the data $D^*\backslash D^*_j$ and then construct
\begin{equation} \label{eq_hatxi}
\hat \bxi_j= \frac{1}{n_j}\sum_{i\in D_j^*} X_iY_i-\frac{1}{n_j}\sum_{i\in D_j^*} X_i\hat h^{-j}(X_i)+\frac{1}{n_j+N_j}\sum_{i\in D_j}X_i\hat h^{-j}(X_i).
\end{equation}
In view of the modified score function (\ref{eq_mscore}), replacing $\bar \bxi$ with $\hat\bxi=(\hat \bxi_1+\hat \bxi_2)/2$, we propose the following semi-supervised Dantzig selector
\begin{equation} \label{est}
\hat\btheta_{SD}=\arg\min \|\btheta\|_1,~~\textrm{s.t.}~~ \|\hat\bSigma_{n+N}\btheta-\hat\bxi\|_\infty \leq \lambda_{SD}.
\end{equation}
Similarly, we define the semi-supervised lasso estimator as
\begin{equation} \label{ssl-lasso}
\hat\btheta_{SL}=\arg\min_{\btheta\in \RR^p} \btheta^T\hat\bSigma_{n+N}\btheta-2\hat\bxi ^T\btheta +2\lambda_{SL}\norm{\btheta}_1.
\end{equation}

\subsection{Theoretical property of the proposed estimator}

We develop the theoretical property for the proposed optimal semi-supervised estimator.

\begin{assumption} \label{assumption_est}
We make the following assumptions:
\begin{enumerate}
\item[(A1)] $\bSigma^{-1/2}X$ is a zero mean sub-Gaussian vector with bounded sub-Gaussian norm and $\Cov(X)=\bSigma$ has smallest eigenvalue $\Lambda_{\min}(\bSigma)\geq C_{\min}>0$ for some positive constant $C_{\min}$. Moreover, $\max_{1\leq j\leq p}\Sigma_{jj}=O(1)$.
\item[(A2)] $\max_{1\leq i\leq n+N}\norm{X_i}_\infty \leq K_1$ where we allow $K_1$ to diverge with $(n,N,p)$.

\item[(A3)] $\EE(\epsilon^2)=\sigma^2$ and {$\EE[(f(X)-X^T\btheta^*)^2]\leq\Phi^2$}. 

\item[(A4)]  $\btheta^*$ is $s$-sparse with $\norm{\btheta^*}_0\leq s$, and $\frac{s\log p}{n+N}=O(1)$.

\end{enumerate}
\end{assumption}

Assumption (A1) is a standard technical condition for $X$ in order to verify the restricted eigenvalue (RE) condition  \citep{bickel2009}. Assumption (A2) imposes the boundedness of the covariates, which simplifies the analysis when the linear model is misspecified \citep{mann2015}. In particular, when $X_i$ is uniformly bounded, $K_1$ becomes a constant. If each component of $X_i$ is Gaussian or sub-Gaussian, Assumption (A2) still holds with high probability with $K_1=C\sqrt{\log [p(n+N)]}$ for some constant $C$. Assumption (A3) only requires the existence of the second moment of $\epsilon$ and $f(X)-X^T\btheta^*$. We note that, {unlike \cite{bickel2009},} we do not assume the residual $Y-X^T\btheta^*$ to be sub-Gaussian. This is because the residual in the misspecified model $Y-X^T\btheta^*=\epsilon+(f(X)-X^T\btheta^*)$ contains the nonlinear term $f(X)-X^T\btheta^*$ which can be large. While we only assume the moment condition in Assumption (A3), the boundedness in Assumption (A2) enables us to apply the Nemirovski moment inequality (Lemma \ref{repeated_lemma}) to control the deviation of the sample estimates from their population. Assumption (A4) is the sparsity condition. In particular, \cite{mann2015} provided some sufficient conditions on $f(X)$ and the distribution of $X$ under which $\btheta^*$ is sparse in the misspecified model.
We further require $\frac{s\log p}{n+N}=O(1)$ to verify the RE condition under the random design; see Lemma \ref{RE}.

Given Assumption \ref{assumption_est}, we establish the convergence rate of the semi-supervised Dantzig selector $\hat \btheta_{SD}$ in (\ref{est}). By \cite{bickel2009},  one can easily show that the same error bounds hold for the semi-supervised lasso estimator $\hat\btheta_{SL}$. For simplicity, we only present the asymptotic results for $\hat \btheta_{SD}$, where $n,p\rightarrow\infty$  and $N$ can be either fixed or tends to infinity as well.

\begin{theorem} \label{mainthm}
Suppose Assumption \ref{assumption_est} holds and the estimator $\hat h^{-j}(\cdot)$ satisfies
$$
\norm{\hat h^{-j}-h}_2=O_p(b_n),
$$
for $j=1,2$, where $b_n$ is a deterministic sequence. Denote $G_h=\norm{h-f}_2$. With some tuning parameter $\lambda_{SD}\asymp K_1(\Phi\sqrt{\frac{\log p}{n+N}}+(\sigma+b_n+G_h)\sqrt{\frac{\log p}{n}})$, the estimator $\hat \btheta_{SD}$ in (\ref{est}) achieves the following error bounds
\begin{equation}\label{eq_mainthm_1}
\norm{\hat \btheta_{SD}-\btheta^*}_q=O_p\Big(
K_1s^{1/q}\Big\{\Phi\sqrt{\frac{\log p}{n+N}}+ (\sigma+b_n+G_h)\sqrt{\frac{\log p}{n}}\Big\}\Big),
\end{equation}
for $q=1, 2$. Moreover,  if $G_h=0$, i.e. $f=h$, $b_n/\sigma=o(1)$ and $K_1=O(1)$, we obtain
\begin{equation}\label{eq_mainthm_2}
\norm{\hat \btheta_{SD}-\btheta^*}_q=O_p\Big(
s^{1/q}\Big\{\Phi\sqrt{\frac{\log p}{n+N}}+ \sigma\sqrt{\frac{\log p}{n}}\Big\}\Big).
\end{equation}
\end{theorem}

{
The error bound (\ref{eq_mainthm_1}) depends on $b_n$, the stochastic error from the estimator $\hat h^{-j}$, and $G_h$ that can be viewed as the asymptotic bias of the estimator.
If we  assume $f$ belongs to $\cF_{\textrm{pairwise}}$ or $\cF_{\textrm{additive}}$ introduced in Section \ref{sec_lower},
there exist estimators $\hat h^{-j}$ in the literature such that $G_h=0$ and $b_n/\sigma=o(1)$; see further explanations in Remark~\ref{rem_consistency} and more detailed examples in Supplement \ref{app_est_f}.
Thus, with $K_1=O(1)$, the error bound (\ref{eq_mainthm_1}) reduces to (\ref{eq_mainthm_2}), which matches the minimax lower bound with $q=1, 2$ in Theorem \ref{thm_lowerbound} up to a $\log s$ factor.
Therefore we call the estimator under this situation the optimal semi-supervised estimator.
On the other hand, if the bias term $G_h$ is large, the error bound (\ref{eq_mainthm_1}) implies that our estimator $\hat \btheta_{SD}$ may have a slow rate. We will revisit this problem in Section \ref{sec_safe}.
}

Practically, we can set the tuning parameter $\lambda_{SD}=CK_1(\hat\Phi\sqrt{\frac{\log p}{n+N}}+\hat\sigma\sqrt{\frac{\log p}{n}})$ and choose the constant $C$ by cross-validation. To account for the magnitude of $\Phi^2$ and $\sigma^2$ in $\lambda_{SD}$, we suggest to estimate $\Phi^2$ and $\sigma^2$ by $\hat \Phi^2=(\hat \Phi_1^2+\hat \Phi_2^2)/2$ and $\hat\sigma^2=(\hat\sigma^2_1+\hat\sigma^2_2)/2$, where $\hat\Phi_j^2=\frac{1}{n_j}\sum_{i\in D_j}(\hat h^{-j}(X_i)-\hat\btheta_{D}^TX_i)^2$ and  $\hat\sigma^2_j=\frac{1}{n_j}\sum_{i\in D^*_j}(Y_i-\hat h^{-j}(X_i))^2$. The cross-validation method works well in our simulations.

\begin{remark}(Examples of $\hat h$).\label{rem_consistency}
If the knowledge of $f(X)$ is available to some extent, we can leverage this information to construct estimators of $f(X)$.
{
We introduced two model classes $\cF_{\textrm{pairwise}}$ and $\cF_{\textrm{additive}}$ in Section \ref{sec_lower}.
For instance, if $f(\cdot)\in \cF_{\textrm{additive}}$, we can directly apply the existing estimators in the literature; see  \cite{lin2006,meier2009,huang2010,additive}, among many others.
}
In particular, Corollary 2 in \cite{huang2010} implies that their adaptive group lasso estimator $\hat h$ constructed with spline basis functions satisfies $\norm{\hat h-f}_2=O_p(n^{-d/(2d+1)})$, where  $d$ is the smoothness of the function $f_k(\cdot)$. 
\end{remark}

\begin{remark}(Comparison with \cite{earliest}).
In the semi-supervised setting, \cite{earliest} proposed a transductive version of lasso and Dantzig selector based on the imputation method.
{
Let $\Tilde {\bY}=(\tilde{Y}_1,...,\tilde{Y}_{n+N})$ denote the imputed outcomes (or pseudo-outcomes) from some preliminary estimator.
}
The transductive Dantzig selector is given by
\begin{equation}\label{eq_trans}
   \hat\btheta_T=\argmin~\norm{\btheta}_1~s.t.~ \frac{1}{n+N}\norm{\tilde\bX(\Tilde{\bY}-\tilde\bX^T\btheta)}_\infty\leq \lambda_T.
\end{equation}
If the imputation satisfies $\frac{1}{n+N}\norm{\tilde\bX(\tilde \bY-\tilde\bX\btheta^*)}_\infty\lesssim (\sigma+\Phi) \sqrt{\frac{\log p}{n}}$, it can be shown that $\|\hat\btheta_T-\btheta^*\|_1\lesssim (\sigma+\Phi)s \sqrt{\frac{\log p}{n}}$ with high probability. We can see that the error bound is of the same order as the supervised Dantzig selector (\ref{eq_ratehat}). Thus, the transductive Dantzig selector is also sub-optimal in the minimax sense; see the last paragraph of Section~\ref{sec_lower}.

To see how the transductive Dantzig selector differs from our estimator $\hat\btheta_{SD}$, we assume that the imputation is attained by using our estimator $\hat h(\cdot)$, i.e., $\tilde Y_i=\hat h(X_i)$ for $1\leq i\leq n+N$.
{
For simplicity, let us omit the cross-fitting step here and defer more derivations in Supplement~\ref{app_hathetaT}.
}
Then the  modified score function $\hat\bSigma_{n+N}\btheta-\hat\bxi$ in our estimator (\ref{est}) can be rewritten as
$$
\hat\bSigma_{n+N}\btheta-\hat\bxi=\frac{1}{n+N}\sum_{i=1}^{n}X_i(Y_i+\frac{N}{n}\{Y_i-\tilde Y_i\}-X_i^T\btheta)+\frac{1}{n+N}\sum_{i=n+1}^{n+N}X_i(\tilde Y_i-X_i^T\btheta).
$$
We can see that for the samples from the unlabeled data ($i\in\{n+1,...,n+N\}$), we use $\tilde Y_i$ as the pseudo-outcome. However, for the samples from the labeled data ($i\in\{1,...,n\}$), $Y_i+\frac{N}{n}\{Y_i-\tilde Y_i\}$ serves as the pseudo-outcome. In comparison, the transductive Dantzig selector (\ref{eq_trans}) always uses $\tilde Y_i$ as the pseudo-outcome for $i\in\{1,...,n+N\}$.


\end{remark}

\begin{remark}(Comparison with \cite{predictionloss}).\label{rem_est_U}
Recently, \cite{predictionloss} proposed a modified lasso estimator for prediction in the semi-supervised setting, which can be reformatted as the following Dantzig selector
\begin{equation} \label{eq_thetaD}
\hat\btheta_{U}=\arg\min \|\btheta\|_1,~~\textrm{s.t.}~~ \|\hat\bSigma_{n+N}\btheta-\frac{1}{n}\sum_{i=1}^{n}X_iY_i \|_\infty \leq \lambda_U,
\end{equation}
where $\hat\bSigma_{n+N}=\frac{1}{n+N}\sum_{i=1}^{n+N}X_i^{\otimes 2}$. Compared with our estimator $\hat\btheta_{SD}$ in (\ref{est}), $\hat\btheta_{U}$ turns out to be a special case of $\hat\btheta_{SD}$ by plugging $\hat h=0$  into (\ref{eq_hatxi}).

\cite{predictionloss}  showed that if a large number of unlabeled data are used to compute $\hat\bSigma_{n+N}$, it becomes more plausible to assume that the compatibility (or RE) constant is bounded away from zero. The statement also holds true for our semi-supervised estimator $\hat \btheta_{SD}$ with the use of $\hat\bSigma_{n+N}$.  Moreover, they proved that the error bound for the excess risk in prediction remains $O_p(s \log p/n)$ under certain conditions, including $|Y|\leq C$ for some constant $C>0$ which indeed implies $\Phi=O(1)$ and $\sigma=O(1)$ by their proof of Theorem 7. To make a fair comparison of $\hat\btheta_{U}$ with our estimator $\hat\btheta_{SD}$, we show that under the same conditions in our Theorem \ref{mainthm},
\begin{equation}\label{eq_ratehatU}
\|\hat\btheta_U-\btheta^*\|_1=O_p\Big(s(\Phi+\sigma+(\btheta^{*T}\bSigma\btheta^*)^{1/2})\sqrt{\frac{\log p}{n}}\Big).
\end{equation}
The proof is deferred to Supplement \ref{app_hathetaU}. It is seen that $\hat\btheta_U$ has a slower rate than our estimator $\hat\btheta_{SD}$ if $\Phi/\sigma\rightarrow\infty$ or $\btheta^{*T}\bSigma\btheta^*/\sigma^2\rightarrow\infty$. Again, in Supplement \ref{app_phi}, we consider the examples where $\Phi^2\asymp s$ and $\sigma$ is a constant, so that $\Phi/\sigma\rightarrow\infty$ holds, as $s$ grows with $n$. Perhaps, a more surprising fact is that the convergence rate of $\hat\btheta_U$ can be even slower than the fully supervised Dantzig selector $\hat\btheta_D$ in (\ref{eq_ratehat}) if $\btheta^{*T}\bSigma\btheta^*/(\sigma^2+\Phi^2)\rightarrow\infty$. Indeed, our simulation studies confirm that the estimator $\hat\btheta_{U}$ often produces larger estimation error than $\hat\btheta_{SD}$ and $\hat\btheta_D$.

\end{remark}

\section{Safe Semi-Supervised Estimator}\label{sec_safe}

Recall from Theorem \ref{mainthm} that our semi-supervised estimator $\hat\btheta_{SD}$ is minimax optimal,
{
if the conditional mean function $f(\cdot)$ can be consistently estimated with a proper rate, e.g., when $f(\cdot)$
belongs to $\cF_{\textrm{pairwise}}$ or $\cF_{\textrm{additive}}$ introduced in Section \ref{sec_lower}.
}
If this does not hold, there is no guarantee that the estimator $\hat\btheta_{SD}$ attains the minimax lower bound or outperforms the supervised estimator. In particular, when $G_h+b_n \gg \Phi$, the convergence rate of $\hat\btheta_{SD}$ can be even slower than the supervised estimator $\hat\btheta_{D}$ in (\ref{dantini}), hence the integration of unlabeled data might fail to improve the estimation accuracy of $\btheta^*$.

To tackle this problem, we develop a safe semi-supervised estimator via a two-step procedure to adapt to the unknown quality of the conditional mean model $h(\cdot)$, which makes the final estimator no worse than the supervised estimator.
To facilitate the theoretical analysis of the two-step estimator, we proceed with the lasso type estimators in this section.
Recall that the supervised lasso was defined in (\ref{eq_lasso}) in Section~\ref{sec_lower}.
Equivalent to (\ref{ssl-lasso}), we can rewrite our optimal semi-supervised lasso estimator $\hat \btheta_{SL}$ as
\begin{equation}\label{est_lassotype}
    \hat \btheta_{SL}=\argmin_{\btheta \in \RR^p} \sum_{j=1}^2\frac{\sum_{i\in D_j}\big(\hat h^{-j}(X_i)-X_i^T\btheta\big)^2}{n+N} -\frac{2\sum_{i\in D_j^*}\big( Y_i-\hat h^{-j}(X_i)\big)X_i^T\btheta}{n}+2\lambda_{SL} \norm{\btheta}_1.
\end{equation}

We construct the safe semi-supervised estimator in the following two steps. First, for a given estimate of the unknown conditional mean $\hat h$, we compute our semi-supervised lasso estimator $\hat \btheta_{SL}$ in (\ref{est_lassotype}) and the supervised lasso estimator $\hat\btheta_L$ in (\ref{eq_lasso}). Since the estimator $\hat \btheta_{SL}$ may not be desirable when the corresponding conditional mean model $h$ is  misspecified, in the second step we further improve the initial estimator $\hat \btheta_{SL}$ by a step of refitting on a suitable estimate of the support set of $\btheta^*$. Denote the support set of two lasso type estimators $\hat \btheta_L$ and $\hat \btheta_{SL}$ by $\hat T_1$ and $\hat T_2$ respectively.
We define the safe semi-supervised estimator as
\begin{equation}\label{safestimator}
    \hat\btheta_R=\hat \btheta_{SL}+\hat \bomega,
\end{equation}
where $\hat \bomega\in\RR^p$ is attained by
\begin{equation}\label{refit}
\hat \bomega=\argmin_{\textrm{supp}(\bomega) \subseteq \hat T_1\cup\hat T_2} \frac{1}{2n}\sum_{i=1}^n(Y_i-X_i^T(\hat \btheta_{SL}+ \bomega))^2+\lambda_\omega\norm{\bomega}_1.
\end{equation}
In  (\ref{refit}) we impose $\textrm{supp}(\bomega) \subseteq \hat T_1\cup\hat T_2$, which implies $\hat \bomega_j=0$ for any $j\notin \hat T_1\cup\hat T_2$. Thus, the safe semi-supervised estimator $\hat\btheta_R$ may only differ from $\hat \btheta_{SL}$ on the estimated support set $\hat T_1 \cup \hat T_2$. Since both the estimands of $\hat \btheta_{SL}$ and $\hat\btheta_R$ are $\btheta^*$, by the definition of (\ref{safestimator}) we can treat $\hat \bomega$ as an estimator of $\btheta^*-\btheta^*=0$. At a first sight, one may expect that estimating a known vector of 0 by $\hat \bomega$ and summing with $\hat \btheta_{SL}$ will inflate the error of the resulting estimator. However, we show a surprising result that the two step estimator $\hat\btheta_R$ may outperform the initial estimator $\hat \btheta_{SL}$.

In the following, we will first explain the intuition behind the estimator $\hat\btheta_R$. Let us consider two scenarios. First, if our initial estimator $\hat \btheta_{SL}$ in step 1 is able to achieve a fast rate, one would expect that $\hat\bomega\approx 0$ due to the $L_1$ regularization in (\ref{refit}). Thus, $\hat\btheta_R$ will inherit the fast rate from the initial estimator $\hat \btheta_{SL}$. In the second scenario, if $\hat \btheta_{SL}$ has a slow rate due to model misspecification, with a suitable choice of tuning parameters, we can obtain $\hat \bomega \approx \hat \btheta_{L}-\hat \btheta_{SL}$. As a toy example, if $p$ is fixed and small and we set all tuning parameters in $\hat \bomega, \hat \btheta_{SL}$ and $\hat \btheta_L$ to be 0, it is easily shown from the least square formula that $\hat \bomega = \hat \btheta_{L}-\hat \btheta_{SL}$. Thus, in this case, the safe semi-supervised estimator will resemble the supervised estimator, since $\hat\btheta_R=\hat \btheta_{SL}+\hat \bomega \approx \hat \btheta_{L}$. In summary, the refitting step can simultaneously retain the fast rate of $\hat \btheta_{SL}$ if it has, and alleviate the potentially unsatisfactory performance of $\hat \btheta_{SL}$ under model misspecification.
The following theorem shows the convergence rate of the safe semi-supervised estimator $\hat \btheta_R$.

\begin{theorem}\label{twostep}
Suppose Assumption \ref{assumption_est} holds, $\Lambda_{\max}(\bSigma)\leq C_{\max}<\infty$ and $s(\log p)^2=o(n)$. Assume that we choose the tuning parameters $\lambda_{SL}$,  $\lambda_L$ and $\lambda_\omega$ in (\ref{est_lassotype}), (\ref{eq_lasso}) and (\ref{refit}) as  $\lambda_{SL}\asymp K_1(\Phi\sqrt{\frac{\log p}{n+N}}+(\sigma+b_n+G_h)\sqrt{\frac{\log p}{n}})$ and $\lambda_L\asymp \lambda_\omega \asymp K_1(\Phi+\sigma)\sqrt{\frac{\log p}{n}}$.  We have for $q=1,2$,  $$\norm{\hat \btheta_R-\btheta^*}_q=O_p(R_{SL}\wedge R_L),$$ where
$
R_{SL}=s^{1/q}K_1\Big(\Phi\sqrt{\frac{\log p}{n+N}}+(\sigma+b_n+\norm{f-h}_2)\sqrt{\frac{\log p}{n}}\Big)$, and $R_L=s^{1/q}K_1(\Phi+\sigma)\sqrt{\frac{\log p}{n}}$.
\end{theorem}

Recall that $R_{SL}$ and $R_L$ correspond to the convergence rates of $\hat\btheta_{SL}$ and $\hat\btheta_L$, respectively. Theorem \ref{twostep} shows that $\hat \btheta_R$ attains the faster rate between $\hat\btheta_{SL}$ and $\hat\btheta_L$. Thus, the estimator $\hat \btheta_R$  remains minimax rate-optimal when the initial estimator $\hat \btheta_{SL}$ is optimal (see Theorem \ref{mainthm}), and is guaranteed to be no worse than the supervised estimators even if the conditional mean model is misspecified.

To choose the tuning parameters, we note that the magnitude of $\lambda_{SL}$ is the same as $\lambda_{SD}$ in Theorem \ref{mainthm}. We can apply the same cross-validation method explained after Theorem \ref{mainthm} to tune $\lambda_{SL}$. Since $\lambda_L$ can be written as $\lambda_L=CK_1\{\EE(Y-X^T\btheta^*)^2\}^{1/2}\sqrt{\frac{\log p}{n}}$ for some constant $C$, one may use scaled lasso to estimate the noise level $\EE(Y-X^T\btheta^*)^2$, and further apply cross-validation to tune $\lambda_L$. In practice, to reduce the computational cost, once we have selected the tuning parameter $\lambda_L$, we can simply set $\lambda_\omega=\lambda_L$, which works well in our simulations.

\begin{remark}
The two-step approach is inspired by the recent works of \cite{proxy} and \cite{transfer} in the context of transfer learning.
However, our theoretical guarantee in Theorem \ref{twostep} is much stronger than those works. Using our terminology, the theoretical analysis in \cite{proxy} and \cite{transfer} implied that the estimator $\hat \btheta_R$ can only attain the worst possible rate between $\hat\btheta_{SL}$ and $\hat\btheta_L$, i.e., $\norm{\hat \btheta_R-\btheta^*}_q=O_p(R_{SL}+R_L)$, which is not sufficient to show $\hat \btheta_R$ to be safe. We obtain a sharper result in Theorem \ref{twostep} because our refitting step (\ref{refit}) differs from those works. In particular, unlike their works, we constrain the support of the estimator $\hat \bomega$ to be $\hat T_1 \cup \hat T_2$, which  guarantees the sparsity of $\hat \btheta_R$.
This is an important intermediate step to prove Theorem \ref{twostep}.
Finally, we note that, in the context of transfer learning, \cite{transfer} also considered how to use model aggregation, such as Q-aggregation \citep{aggregation}, to improve the initial estimator.
We refer to Supplement~\ref{counter} for further discussion on model aggregation.

\end{remark}

\section{Aggregation of Semi-Supervised Estimators} \label{sec_safe2}

Recall from Theorem~\ref{mainthm} that the performance of the semi-supervised estimator depends on the estimator $\hat h$ of the conditional mean model.
In practice, it is uncommon for us to find a consistent estimator of the conditional mean function.
More commonly, we may face multiple choices of $\hat h$'s,  none of which is consistent.
We may expect that different $\hat h$'s only capture different aspects of the non-linearity of the conditional mean function.
Does aggregating multiple semi-supervised estimators help us explore the non-linearity of the conditional mean function? Here, we generalize the two-step method in Section \ref{sec_safe} to answer this question.

Assume that two different estimators of the conditional mean function $\hat h_1$ and $\hat h_2$  are available. We denote by $\hat \btheta_{h_1}$ and $\hat \btheta_{h_2}$ the semi-supervised lasso estimators in (\ref{est_lassotype}) with $\hat h_1$ and $\hat h_2$ and tuning parameters $\lambda_{h_1}$ and $\lambda_{h_2}$ respectively. In the following, we will apply the two-step procedure to combine $\hat \btheta_{h_1}$ and $\hat \btheta_{h_2}$. We first
compute $\hat \btheta_{h_1}$ and $\hat \btheta_{h_2}$ in step 1. Their support sets are denoted by $\hat H_1$ and $\hat H_2$. In step 2, we use $\hat \btheta_{h_1}$ as the initial estimator and define the aggregated estimator as
\begin{equation}\label{est_ag}
    \hat \btheta_{AH}=\hat \btheta_{h_1}+\hat \bomega_h,
\end{equation}
where
 \begin{align*}
    \hat \bomega_h=\argmin_{\textrm{supp}(\bomega)\subseteq \hat H_1\cup \hat H_2} &\sum_{j=1}^2\Big(\frac{\sum_{i\in D_j}\big(\hat h_2^{-j}(X_i)-X_i^T(\hat \btheta_{h_1}+\bomega)\big)^2}{n+N} \\
    & ~~~~~~~~~ -\frac{2\sum_{i\in D_j^*}\big( Y_i-\hat h_2^{-j}(X_i)\big)X_i^T(\hat \btheta_{h_1}+\bomega)}{n}\Big)
    +2\lambda_H \norm{\bomega}_1.
\end{align*}
Unlike the refitting step in (\ref{refit}), we also incorporate the unlabeled data to compute $\hat \bomega_h$
so that $\hat \btheta_{AH}$ can attain a better convergence rate than $\hat \btheta_{h_1}$ or $\hat \btheta_{h_2}$.
One may note that the creation of $\hat \btheta_{AH}$ is not symmetric to $h_1$ and $h_2$; however, our theoretical analysis below shows that the priority order of $h_1$ and $h_2$ does not really matter.

\begin{proposition}\label{twostep_h}
Suppose Assumption \ref{assumption_est} holds, $\Lambda_{\max}(\bSigma)\leq C_{\max}<\infty$ and $s(\log p)^2=o(n+N)$. The estimators $\hat h_1$ and $\hat h_2$ satisfy $\|\hat h_1^{-j}-h_1\|_2=O_p(b_{h_1})$ and $\|\hat h_2^{-j}-h_2\|_2=O_p(b_{h_2})$ for $j=1,2$. Denote $G_{h_1}=\|h_1-f\|_2$ and $G_{h_2}=\|h_2-f\|_2$. Selecting some tuning parameters $\lambda_{h_1}\asymp K_1(\Phi\sqrt{\frac{\log p}{n+N}}+(\sigma+b_{h_1}+G_{h_1})\sqrt{\frac{\log p}{n}})$,  $\lambda_{h_2}\asymp\lambda_H \asymp K_1(\Phi\sqrt{\frac{\log p}{n+N}}+(\sigma+b_{h_2}+G_{h_2})\sqrt{\frac{\log p}{n}})$, we can show that for $q=1,2$,
$$\norm{\hat \btheta_{AH}-\btheta^*}_q=O_p(R_{h_1}\wedge R_{h_2}),$$ where
$R_{h_j}=s^{1/q}K_1\Big(\Phi\sqrt{\frac{\log p}{n+N}}+(\sigma+b_{h_j}+\norm{f-h_j}_2)\sqrt{\frac{\log p}{n}}\Big)$.
\end{proposition}

This proposition shows that the aggregated estimator $\hat \btheta_{AH}$ attains the best possible rate between $\hat \btheta_{h_1}$ and $\hat \btheta_{h_2}$. Following the same reasoning, one may use $\hat \btheta_{AH}$ as the initial estimator and repeat the refitting step, if a third estimator $\hat \btheta_{h_3}$ is available. The resulting estimator attains the best possible rate among the three estimators $\hat \btheta_{h_1}$,  $\hat \btheta_{h_2}$ and $\hat \btheta_{h_3}$. We expect that, in general, our two-step procedure can be applied to aggregate  multiple estimators, as long as the number of the candidates is fixed and small. We refer to Section \ref{sec_simu} for numerical performance of the aggregated estimators.

\section{Simulation Studies}\label{sec_simu}

\subsection{Data generating models and practical implementation}
We first generate a $p$-dimensional multivariate normal random vector $Z \sim \mathcal{N}(0,\Sigma)$ with  $\Sigma_{jk}=0.3^{|j-k|}$. We set the covariate $X=(X_1,...,X_p)$ to be $X_1=|Z_1|$ and $X_j=Z_j ~\mathrm{for}~ 1<j\leq p$. 
The reason we take $X_1=|Z_1|$ is that this transformation implies $\EE(X_1^k X_j)=0$ for $j\neq 1$ but the parameter $\theta^*_1$ for centered $X_1$ is nonzero. We consider the following three data generating models for $Y$. For Model 1, we consider an additive model
$$
Y=0.5X_1^2+0.8X_3^3-(X_4-2)^2+2(X_5+1)^2+2X_6+\epsilon,
$$
where $\epsilon \sim \mathcal{N}(0,1)$.
To calculate the corresponding regression parameter $\btheta^*$ under the working linear model, we first center $Y$ and $X_1$ so that their means are 0. By Proposition 4 in \cite{mann2015}, we know that the support of $\btheta^*$ is $S=\{1,3,4,5,6\}$ and $\theta_j^*$ for any $j\in S$ is given by the $L_2(\PP)$ projection in the sub-model only with the variable $X_j$ (e.g, $\theta_3^*=\arg\min \EE(0.8X_3^2-\theta_3 X_3)^2$). After some calculation, we obtain $\btheta^*=(1.1,0,2.4,4,4,2,0,...,0)$, which is sparse.

For Model 2, we consider
$$Y=0.6(X_1+X_2)^2+0.4X_4^3-X_5+2X_6+\epsilon,$$
where $\epsilon \sim \mathcal{N}(0,1)$. The model is non-additive since it includes an interaction term between $X_1$ and $X_2$.
The corresponding regression parameter $\btheta^*$ is $(1.48,1.04,0,1.2,-1,2,0,...,0)$.

Besides, we consider a non-additive Model 3 which includes extra randomly selected support sets $S_1$ and $S_2$  on top of Model 1. We choose $|S_1|=|S_2|=5$ and set
$$
Y=0.5X_1^2+0.8X_3^3-(X_4-2)^2+2(X_5+1)^2+2X_6+2\sum_{k\in S_1}X_k+0.05(\sum_{k\in S_2}X_k)^3+\epsilon,
$$
where $\epsilon \sim \mathcal{N}(0,1)$. In this model, $\theta^*_k=0.75$ for $k\in S_2$, $\theta^*_k=2$ for $k\in S_1$, and the other components of $\btheta^*$ are the same as in Model 1. Under each data generating model, we consider several combinations of $(n,p)$ and vary the ratio $N/n$ in a certain range. We repeat the simulation 100 times and output the average as final results.

Before we proceed to illustrate the results, we list the estimators considered in this section and discuss several practical implementation issues. Since the performance of lasso and Dantzig type estimators are similar, we only consider the Dantzig type estimators here for simplicity.
\begin{itemize}
    \item The proposed semi-supervised estimator $\hat\btheta_{SD}$ in (\ref{est}) (SSL1) with a sparse additive model $h_1$. We estimate the sparse additive model by using the group lasso with the spline basis \citep{huang2010}. To be specific, we use the cubic spline basis with degree of freedom $df=5$. To select the penalty parameter in group lasso and make computation easier, the BIC criterion is used; see Section 4 in \cite{huang2010} for the definition. After we derive the estimator $\hat h_1$ and subsequently $\hat \bxi$, we modify the source code in the \textsf{flare} package to compute $\hat\btheta_{SD}$, where the tuning parameter $\lambda_{SD}$ is selected by 5 fold cross-validation.
    \item The proposed semi-supervised estimator $\hat\btheta_{SD}$ in (\ref{est}) (SSL2) with a pairwise interaction model $h_2$. The model $h_2(\cdot)$ corresponds to the linear regression containing all the linear terms, the squares of the variable and the interaction terms \citep{zhao2016analysis}.
    \item The supervised Dantzig selector $\hat\btheta_D$ in (\ref{dantini}) (Dantzig). We use the \textsf{flare} package to compute the estimator and select the tuning parameter by 5 fold cross-validation.
    \item The modified Dantzig selector $\hat \btheta_U$ in (\ref{eq_thetaD}) (U-Dantzig).
    \item The aggregated estimator $\hat\btheta_{AH}$ in (\ref{est_ag}) (SSL12) that combines SSL1 and SSL2.
    \item The safe semi-supervised estimator $\hat\btheta_R$ (S-SSL1) using SSL1 as the initial estimator. As seen from the discussion after Theorem \ref{twostep}, we set the tuning parameter $\lambda_\omega$ identical to the tuning parameter for the supervised Dantzig selector.
    \item The safe semi-supervised estimator $\hat\btheta_R$ (S-SSL12) using SSL12 (e.g., the aggregated estimator) as the initial estimator.
\end{itemize}

\subsection{Numerical results}

\begin{figure}
    \centering
    \includegraphics[scale=0.55]{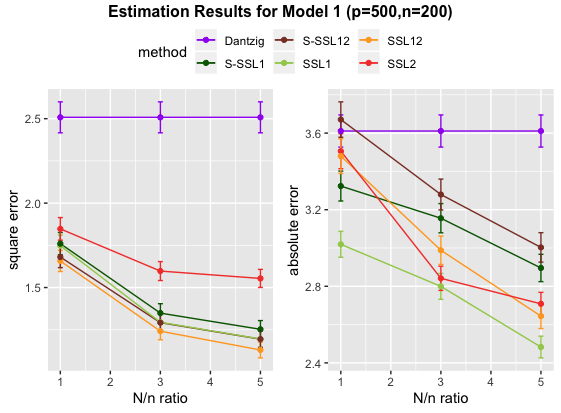}
    \caption{The $L_2$ and $L_1$ estimation errors under Model 1 with $p=500$ and $n=200$. The length of the vertical bar represents the magnitude of the sample standard deviations. $L_2$ errors for U-Dantzig are 5.75 (0.20), 7.14 (0.24), 7.47 (0.26) and $L_1$ errors for U-Dantzig are 6.15 (0.17), 6.53 (0.15), 6.73 (0.15). The numbers in the parenthesis are sample standard deviations.}
    \label{Fig:EM1p5}
\end{figure}

Under Model 1 with $p=500$ and $n=200$, the comparisons of the estimation errors of SSL1, SSL2, Dantzig, SSL12, S-SSL1 and S-SSL12 are illustrated in Figure \ref{Fig:EM1p5}, while the results from U-Dantzig are shown in numbers in the caption of the figure due to the scale of plot.
Since the true data generating model is additive with some quadratic terms, both SSL1 and SSL2 can leverage the non-linearity of the conditional mean function and their estimation errors are much smaller than Dantzig, which agrees with Theorem \ref{mainthm}. Besides, we can see that SSL1 outperforms SSL2, since the imposed additive model in SSL1 can better estimate the conditional mean function.  By aggregating SSL1 and SSL2, our estimator SSL12 achieves the minimum $L_2$ error among all those methods. For the two safe semi-supervised estimators (S-SSL1 and S-SSL12), they retain the optimal rate in $L_2$ norm from the corresponding semi-supervised estimators (SSL1 and SSL12), and clearly outperform Dantzig, which is consistent with the theoretical property in Theorem \ref{twostep}.


One interesting observation is that U-Dantzig performs much worse than the fully supervised estimator Dantzig; see the caption of Figure \ref{Fig:EM1p5}. Thus, using the sample covariance $\hat \bSigma_{n+N}$ from both labeled and unlabeled data in the Dantzig selector may not provide any empirical improvement; see Remark \ref{rem_est_U} for the theoretical justification.
In addition, as the size of unlabeled data $N$ increases, the improvement of our semi-supervised estimators (SSL1, SSL2, SSL12, S-SSL1 and S-SSL12) is more overwhelming, whereas the performance of U-Dantzig tends to deteriorate.

\begin{figure}
    \centering
    \includegraphics[scale=0.55]{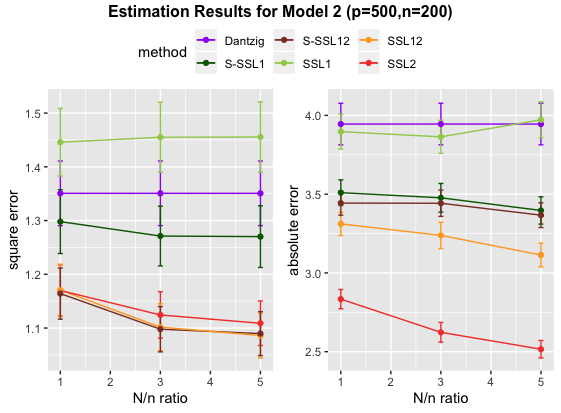}
    \caption{The $L_2$ and $L_1$ estimation errors under Model 2 with $p=500$ and $n=200$. The length of the vertical bar represents the magnitude of the sample standard deviations. $L_2$ errors for U-Dantzig are 2.31 (0.08), 2.75 (0.09), 2.87 (0.11) and $L_1$ errors for U-Dantzig are 4.28 (0.10), 4.62 (0.10), 4.66 (0.07).}
    \label{Fig:EM2p5n2}
\end{figure}

The comparisons under Model 2 with $p=500$ and $n=200$ are shown in Figure \ref{Fig:EM2p5n2}. Since Model 2 includes an interaction term between $X_1$ and $X_2$, the sparse additive model $h_1$ is inconsistent for the true regression function. Thus SSL1 does not improve the estimation accuracy compared with fully supervised Dantzig. However, the safe semi-supervised estimator, S-SSL1, successfully mitigates the undesired performance of SSL1 and its $L_1$ and $L_2$ errors are smaller than Dantzig. This agrees with Theorem \ref{twostep} that the refitting step provides a safe use of unlabeled data even if the imposed conditional mean model is incorrect. On the other hand, since the pairwise interaction model $h_2$ is a correctly specified conditional mean model, the estimators SSL2, SSL12 and S-SSL12, that depend on this model, show small estimation errors.

\begin{figure}
    \centering
    \includegraphics[scale=0.55]{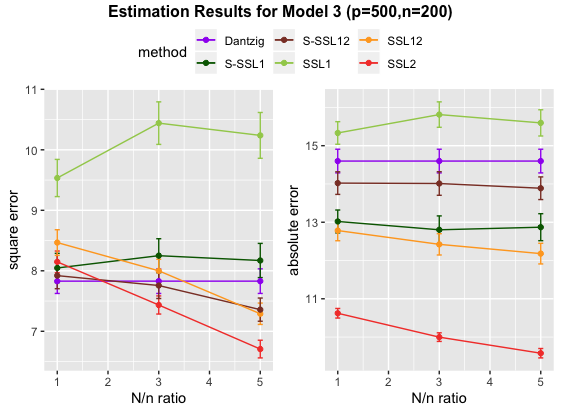}
    \caption{The $L_2$ and $L_1$ estimation error under Model 3 with $p=500$ and $n=200$. The length of the vertical bar represents the magnitude of the sample standard deviations.
    $L_2$ errors for U-Dantzig are 11.46(0.23), 13.29(0.25), 13.62(0.26) and $L_1$ errors for U-Dantzig are 14.32(0.16), 15.73(0.24), 15.60(0.20).}
    \label{Fig:EM3p5}
\end{figure}

Under Model 3, since the true conditional mean function differs significantly from the additive model $h_1$, SSL1 yields large estimation errors. Nevertheless, the performance of S-SSL1 is comparable and no worse than the fully supervised Dantzig. While the pairwise interaction model $h_2$ cannot account for the third order interaction terms in the set $S_2$, it can still partially explain the non-linearity of the true conditional mean function. Thus, the performance of SSL2 is still better than Dantzig in $L_2$ norm when $N/n=5$ and also in $L_1$ norm.
{
The comparison results are summarized in Figure~\ref{Fig:EM3p5}.
}

{
The Supplement \ref{app:plots} contains further simulation results with $p=200$ and $n=100$, and with $p=1000$ and $n=300$, for all the three models considered above, as well as some other numerical results.
}

\section{Real Data Application}\label{sec_data}

\begin{figure}
    \centering
    \includegraphics[scale=0.55]{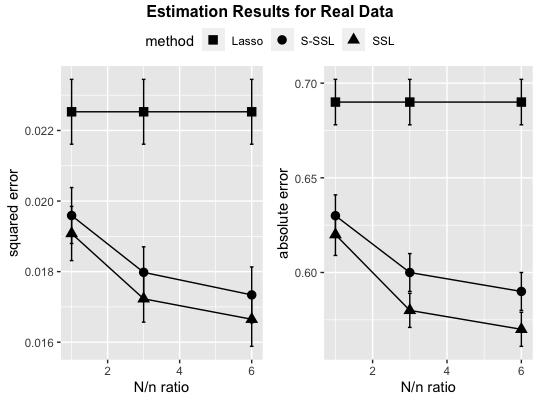}
    \caption{The $L_2$ and $L_1$ estimation error for real data application. The length of the vertical bar represents the magnitude of the sample standard deviations.}
    \label{Fig:real data}
\end{figure}

In this section, we illustrate our proposed methods in a real data example, derived from the Medical Information Mart for Intensive Care III (MIMIC-III) database \citep{johnson2016mimic}. MIMIC-III is an openly available electronic health records system developed by the MIT Lab for Computational Physiology. It contains de-identified health-related data for 38,597 adult patients (aged 16 years or above) admitted to intensive care units (ICU) of the Beth Israel Deaconess Medical Center between 2001 and 2012.
Some details of the adult patients by their first ICU admissions are available in the Table 1 of \cite{johnson2016mimic}.

Researchers have used the MIMIC-III database to investigate a variety of medical issues such as predicting ICU readmissions \citep{brown2012epidemiology, tabak2017predicting, xue2019predicting} and associating various clinical biomarkers with mortality \citep{liu2020serum, du2021prediction, jhou2021plasma, tang2021association}.
Our initial motivation for this data analysis is to understand the relation between the albumin level in the blood sample, oftentimes used to screen for liver or kidney disease \citep{phillips1989association}, and all other variables including demographics, chart events, and clinical biomarkers from the laboratory tests.

After all of the data pre-processing steps detailed in Supplement~\ref{app:plotsdata}, we are left with a dataset with 4784 patients and each of them has $p=2928$ covariates.
The results depicted in Figure~\ref{Fig:real data} are an aggregation from 100 replications.
In each replication, we randomly choose 3500 patients and call the first $n=500$ patients the labeled data. We mask the outcome ``albumin'' for all other 3000 patients and call them the unlabeled data. In each experiment, we gradually increase the sample size of the unlabeled data from $N=500$ to $N=1500$ and to $N=3000$.
{
Since the working model already includes some pairwise interaction terms among biomarkers, we use the random forest to estimate the conditional mean function in our SSL estimator and the corresponding S-SSL estimator.
}
{
The $L_1$ and $L_2$ estimation errors presented in Figure 4 are computed against the lasso estimator using all of the 4784 samples, which is regarded as the proxy of the underlying true linear coefficients of the working model.
}

From Figure~\ref{Fig:real data}, as the ratio $N/n$ increases from 1 to 6, compared to supervised lasso, the proposed S-SSL estimator could bring the $L_1$ error down around 8.7\% to 14.5\% and the proposed SSL estimator could bring the $L_1$ error down around 10.1\% to 17.4\%, respectively.
The percentage decrease for the $L_2$ error ranges from 13.0\% to 23.0\% for the S-SSL estimator and from 15.3\% to 26.1\% for the SSL estimator.
Compared to the S-SSL estimator, the outperformance of the SSL estimator, albeit not as significant as the comparison with the supervised lasso, is still noticeable from Figure~\ref{Fig:real data}.
{
The likely reason for this is that the random forest used in our semi-supervised estimators (SSL and S-SSL) has already effectively captured the structure of the conditional mean function. Thus, S-SSL behaves similarly to SSL.
}
All of these observations support the effectiveness of the methods proposed in this paper.

{
Finally, Supplement~\ref{app:plotsdata} contains the detailed data cleaning and data pre-processing procedures for this application, as well as some other results and conclusions.
Along the paper, we also submit the programming code for anyone who has interest to reproduce the results.

}






\section*{Acknowledgment}

Ning is supported in part by U.S. National Science Foundation (NSF, DMS 1941945 and DMS 1854637). Zhao is supported in part by NSF (DMS 2122074). Zhang is supported in part by U.S. National Institutes of Health (R01HG010171 and R01MH116527) and NSF (DMS 2112711).
The authors would like to thank the Editor, an Associate Editor, and three reviewers for their insightful comments which have helped improve the manuscript substantially.

\bibliographystyle{asa}
\bibliography{reference}

\end{document}


\setcounter{page}{26} 
\title{Supplemental Materials for ``Optimal and Safe Estimation for High-Dimensional Semi-Supervised Learning''}

\maketitle




\section*{Supplementary Materials}

\setcounter{equation}{0}\renewcommand{\theequation}{S.\arabic{equation}}
\setcounter{subsection}{0}\renewcommand{\thesubsection}{S.\arabic{subsection}}

\newtheorem{lem}{{\bf Lemma}}
\setcounter{lem}{0}\renewcommand{\thelem}{S.\arabic{lem}}

\newtheorem{defi}{{\bf Definition}}
\setcounter{defi}{0}\renewcommand{\thedefi}{S.\arabic{defi}}

\subsection{Sub-Gaussian variable and vector}

To characterize the tail behavior of random variables, we introduce the following definition.

\begin{defi}[Sub-Gaussian variable and vector]
A random variable $X$ is called sub-Gaussian if there exists a positive constant $K_2$ such that $\PP(|X|>t)\leq \exp(1-t^2/K^2_2)$ for all $t\geq 0$. The sub-Gaussian norm of $X$ is defined as $\|X\|_{\psi_2}=\sup_{p\geq 1}p^{-1/2}(\EE|X|^p)^{1/p}$. A vector $\bX\in\RR^p$ is a sub-Gaussian vector if the one-dimensional marginals $\bv^T\bX$ are sub-Gaussian for all $\bv\in\RR^p$, and its sub-Gaussian norm is defined as $\|\bX\|_{\psi_2}=\sup_{\|\bv\|_2=1}\|\bv^T\bX\|_{\psi_2}$.
\end{defi}

\subsection{Illustration of Convergence Rate}\label{sec:rateplot}

\begin{figure}[htbp]
\centering
\subfigure{\includegraphics[scale=0.6]{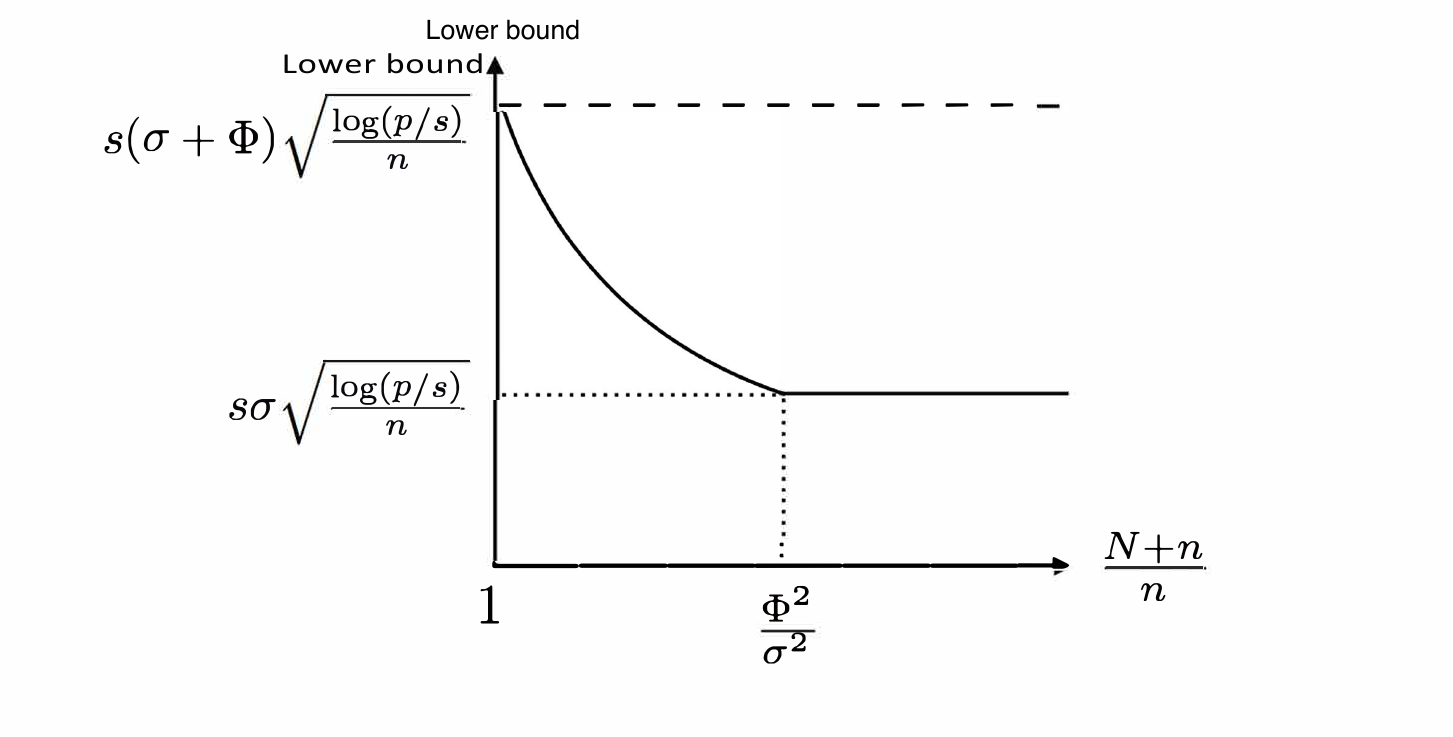}}
   \vspace{-6mm}
  \caption{Plot of the lower bound with $q=1$ in Theorem \ref{thm_lowerbound} (the solid curve) and the upper bound from Dantzig selector in (\ref{eq_ratehat}) (the dashed line), against the value of $(N+n)/n$. The region between the two corresponds to the gap between the lower bound for semi-supervised estimators and the upper bound obtained from the supervised estimators. In this plot, we fix $n$ and vary the value of $N$.  \label{fig_bound}}
 \end{figure}

\subsection{Proofs}\label{app_mainproof}

Throughout the proofs without causing extra confusions, for notation simplicity, we use $C, C', c_1, c_2$ etc. to denote generic constants whose values can change from time to time.

\subsubsection{Preliminary Lemmas}
We start with several basic lemmas that we will apply in our proofs.

\begin{lem}[Lemma B.1 in \cite{victor}] \label{chern}
 Let $\{X_n\},\{Y_n\}$ be sequences of random variables. If for any $c>0$, $\PP(|X_n|>c|Y_n)=o_p(1)$. Then $X_n=o_p(1)$.
 \end{lem}

\begin{lem}[Nemirovski moment inequality, Lemma 14.24 in \cite{shd}]\label{repeated_lemma}
For $m\geq1$ and $p>e^{m-1}$, we have \begin{equation}
    \EE\Big[\max_{1\leq k\leq p}\Big|\sum_{i=1}^n(\gamma_k(Z_i)-\EE[\gamma_k(Z_i)])\Big|^m\Big]\leq (8\log 2p)^{\frac{m}{2}}\EE\Big[(\max_{1\leq k\leq p}\sum_{i=1}^n\gamma_k^2(Z_i))^{m/2}\Big]
\end{equation}

\end{lem}

\begin{lem}[Theorem 3.1 in \cite{RE}]\label{RE}
 Assume that $\bX\in \RR^{n\times p}$ has zero mean and covariance $\bSigma$. Furthermore, assume that the rows of $\bX\bSigma^{-1/2}\in \RR^{n\times p}$ are independent sub-gaussian random vector with a bounded sub-gaussian constant and $\Lambda_{\min}(\bSigma)>C_{\min}>0$,  $\max_{1\leq j\leq p}\Sigma_{jj}=O(1)$. Set $0<\delta<1$, $0<s_0<p$, and $L>0$.  Define the following event,
 \begin{equation} \label{lowerRE}
     \mathcal{B}_\delta(n,s_0,L)=\{\bX\in \RR^{n\times p}:(1-\delta)\sqrt{C_{\min}} \leq \frac{\norm{\bX \bv}_2}{\sqrt{n}\norm{\bv}_2},\forall \bv\in \mathcal{C}(s_0,L)~\textrm{s.t.}~\bv\neq 0\}.
 \end{equation}
and $\mathcal{C}(s_0,L)=
\{\btheta\in \RR^p:\exists S\subseteq \{1,...,p\},|S|=s_0, \norm{\btheta_{S^c}}_1\leq L\norm{\btheta_S}_1\}$. Then, there exists a constant $c
_1=c(L,\delta)$ such that, for sample size $n\geq c_1s_0\log(p/s_0)$, we have
\begin{equation}
\PP(\mathcal{B}_\delta(n,s_0,L))\geq 1-e^{-\delta^2n}.
\end{equation}

\end{lem}

\begin{lem}\label{bdant}
Assume that  Assumption \ref{assumption_est} holds. Consider the Dantzig selector $\hat \btheta_D$ in (\ref{dantini}) with $\lambda_D\asymp K_1\sqrt{\frac{(\sigma^2+\Phi^2)\log p}{n}}$. We have
\begin{equation}\label{eq_bdant1}
\norm{\hat \btheta_D-\btheta^*}_1=O_p(s\lambda_D),~~\textrm{and}~~\frac{1}{n}\sum_{i=1}^n[X_i^T(\hat \btheta_D-\btheta^*)]^2=O_p(s\lambda_D^2).
\end{equation}
Moreover, we have
$$
\|\hat\bSigma_n(\hat \btheta_D-\btheta^*)\|_{\infty}=O_p(\lambda_D),
$$
where $\hat\bSigma_n=\frac{1}{n}\sum_{i=1}^nX_i^{\otimes 2}$.
\end{lem}

\begin{proof}
The proof of the convergence rate of $\hat\btheta_D$ in (\ref{eq_bdant1}) is similar to Theorem 7.1 in \cite{bickel2009}. The key step is to derive
$$
\norm{\frac{1}{n}\sum_{i=1}^n X_i(Y_i-X_i^T\btheta^*)}_\infty\lesssim K_1(\sigma^2+\Phi^2)^{1/2}\sqrt{\frac{\log p}{n}},
$$
which is implied by Lemma \ref{repeated_lemma} together with $\EE(Y_i-X_i^T\btheta^*)^2=\sigma^2+\Phi^2$ and $\|X_i\|_\infty\leq K_1$. The rest of the proof is omitted. To show the rate of $\|\hat\bSigma_n(\hat \btheta_D-\btheta^*)\|_{\infty}$, we note that,  with $\lambda_D=C K_1\sqrt{\frac{(\sigma^2+\Phi^2)\log p}{n}}$ for some sufficiently large $C$, we have
\begin{align*}
\|\hat\bSigma_n(\hat \btheta_D-\btheta^*)\|_{\infty}&\leq \|\frac{1}{n}\sum_{i=1}^n X_i(Y_i-X_i^T\hat\btheta_D)\|_{\infty}+\|\frac{1}{n}\sum_{i=1}^n X_i(Y_i-X_i^T\btheta^*)\|_{\infty}\\
\leq 2\lambda_D,
\end{align*}
where we invoke the constraint of $\hat\btheta_D$ as the definition of Dantzig selector in the last step.
\end{proof}

{
\subsubsection{Proof of Theorem \ref{thm_lowerbound}}
To simplify the notation, we use $\mathcal{P}_{\Phi,\sigma}$ to denote $\mathcal{P}^{\textrm{pairwise}}_{\Phi,\sigma}$ or $\mathcal{P}^{\textrm{additive}}_{\Phi,\sigma}$ in the proof. The proof is very similar under these two classes, and we will point out the difference when it occurs.

The lower bound consists of two parts. First, by taking $f(X)=X^T\btheta^*$, that is when the linear model is correctly specified, the proof in Proposition 6.4 of \cite{verz} with conditions $s\log(p/s)\leq C'n$ and $2\leq s\leq (n-1)/4$ directly implies
$$
\inf_{\hat\theta} \sup_{P_{X,Y}\in \mathcal{P}_{\Phi,\sigma}} \PP_{P_{X,Y}} \Big[\norm{\hat\btheta-\btheta^*}_q\gtrsim c_1s^{1/q}\sigma\sqrt{\frac{\log (p/s)}{n}}\Big]>c_2.
$$
We note that taking $f(X)=X^T\btheta^*$ is feasible under $\mathcal{P}^{\textrm{pairwise}}_{\Phi,\sigma}$ and $\mathcal{P}^{\textrm{additive}}_{\Phi,\sigma}$, and we also use the assumption $\bar s\geq s$.

Second, to establish the lower bound $s^{1/q}\Phi\sqrt{\log (p/s)/(n+N)}$, we first construct a set of hypotheses and then apply Theorem 2.7 in \cite{tsy}. In this case, we take $f(X)=b f_u(X_u)$, for some $u\subseteq \{1,2,...,p\}$ with $a= |u|\leq s/2$ and $b\in \RR$ to be set later. For example, under $\mathcal{P}^{\textrm{pairwise}}_{\Phi,\sigma}$, we can set $f_u(X_u)=X_1X_2$; under $\mathcal{P}^{\textrm{additive}}_{\Phi,\sigma}$, we can set $f_u(X_u)=X_1^2$ (note that $x^2$ is a second order smooth function). Without the loss of generality, we set $u=\{1,...,a\}$.

Define the set $\mathcal{M}=\{x\in \{0,1\}^{p-a}: \|x\|_0=s/2\}$. It follows from the Varshamov-Gilbert bound (e.g., Lemma 2.9 in \cite{tsy}) and Lemma A.3 in \cite{rigollet2011} that there exists a subset $\mathcal{M}'$ of $\mathcal{M}$ such that for any $x,x'$ in $\mathcal{M}'$ with $x\neq x'$, we have
\begin{equation}\label{eq_thm_lowerbound_0}
\rho_H(x,x')>\frac{s}{16},~~\textrm{and}~~\log|\mathcal{M}'|\geq c_1's\log(\frac{p-a}{s}),
\end{equation}
where $\rho_H$ denotes the Hamming distance and $c_1'>0$ is an absolute constant. Denote the element of the finite set $\mathcal{M'}$ by $\bw^j$ and the index set for the nonzero entries by $[j]$. For each $\bw^j\in\RR^{p-a}$, we extend the vector by adding zero at the first $a$ entries, i.e. the corresponding element in $u$ to obtain a $p$-dimensional vector $(0_a,\bw^j)$, and for notational simplicity, we still call it $\bw^j$.

Next, we construct a finite set of hypotheses by perturbing the distribution of $X$. Denote $N=(N_1,...,N_p)\sim \mathcal{N}(0,I_p)$. Let us consider the following hypothesis
$$\begin{aligned}
     H_0&: X=(X_1,...,X_p) ~\textrm{with}~ X_{\ell}=N_{\ell},
\end{aligned}$$
and under $H_0$, denote $\btheta^*= \arg \min_{\btheta \in \mathbb{R}^p}\EE_0[(f(X)-X^T\btheta)^2]$. It is easily seen that $\btheta^*_{u^c}=0$. With $f^{\perp}(X) = f(X)-X^T\btheta^*$, we can choose $b$ so that $\EE_0[(f^\perp(X))^2]=\Phi^2$. We construct other hypotheses as
$$\begin{aligned}
     H_j&: X=(X_1,...,X_p) ~\textrm{with}~ X_{\ell}=
     \begin{cases}
     \frac{\rho}{\Phi} f^{\perp}(N)+ \sqrt{1-\rho^2}N_{\ell}, &\mbox{ if } \bw^j_\ell=1,\\
     N_\ell, &\mbox{ if } \bw^j_\ell=0,
     \end{cases}
\end{aligned}$$
for $j=1,...,|\cM'|$, where $\rho>0$ is a quantity to be chosen later. 
Let $\EE_j$ denote the expectation under $H_j$. Clearly, $\EE_j(X_\ell)=0$. After some simple calculation, we can verify that for $j=1,...,|\cM'|$,
\begin{equation}\label{eq_thm_lowerbound_1}
\EE_j[X_\ell X_m]=\begin{cases}
\rho^2 & \mbox{if}~ \bw^j_\ell ~\mbox{and}~ \bw^j_m \neq 0 ~\mbox{and}~ \ell\neq m, \\
1 & \mbox{if}~ \ell=m,\\
0 & \mbox{otherwise.}
\end{cases}
\end{equation}
Denote by $\Mb=\EE_j[XX^T]$ the covariance matrix of $X$ under $H_j$. From (\ref{eq_thm_lowerbound_1}), there exists a permutation matrix $\Pb$ such that $\Mb=\Pb\Bb\Pb^T$, where $\Bb=\textrm{diag}(\Ab,I_{p-\frac{s}{2}})$ and $\Ab$ is a $s/2$-dimensional equicorrelation matrix with the off-diagonal entry $\rho^2$ and the diagonal entry 1. Assume that the $(\lambda,\bv)$ are the eigenvalue and corresponding eigenvector of $\Bb$. Following the definition of eigenvalues, $(\lambda,\Pb\bv)$ are the eigenvalue and eigenvector of $\Mb$. As a result, $\lambda_{\min}(\Mb)=\lambda_{\min}(\Bb)=\lambda_{\min}(\Ab)=1-\rho^2$, where the last two equalities follow from the property of the block diagonal matrix and equicorrelation matrix. With the choice of $\rho$ as specified in (\ref{eq_thm_lowerbound_rho}), we can derive that $\lambda_{\min}(\Mb)>1/2$ and therefore $\Mb$ is positive definite.

Our next step is to verify that the corresponding estimand $\btheta^j=\{\EE_j(XX^T)\}^{-1}\EE_j(Xf(X))$ is $s$-sparse. To this end, we first note that \begin{equation*}
    \EE_j[X_{\ell} f(X)]=\begin{cases}
    \rho\Phi, &\mbox{ if } \bw^j_\ell=1,\\
     \EE_j[N_lf(N)] = \btheta^*_l, &\mbox{ if } \bw^j_\ell=0.
    \end{cases}
\end{equation*}
To calculate $\Mb^{-1}$, we first rewrite the equicorrelation matrix $\Ab$ as $\Ab=(1-\rho^2)I_{\frac{s}{2}}+ \rho^2\bm{1}_{\frac{s}{2}}\bm{1}_{\frac{s}{2}}^T$, where $\bm{1}_{\frac{s}{2}}$ is a ${\frac{s}{2}}$-dimensional vector of $1$. The Woodbury formula implies
$$
\Ab^{-1}=\frac{1}{1-\rho^2}I_{\frac{s}{2}}-\frac{\rho^2}{(1-\rho^2)(1+({\frac{s}{2}}-1)\rho^2)}\bm{1}_{\frac{s}{2}}\bm{1}_{\frac{s}{2}}^T.
$$
Thus, we obtain $\Mb^{-1}=\Pb\Bb^{-1}\Pb^{T}$ with $\Bb^{-1}=\textrm{diag}(\Ab^{-1}, I_{p-\frac{s}{2}})$. Let $\be_t$ denote the canonical basis in $\RR^p$ with the $t$-th entry being 1 and the rest being 0. Note that the permutation matrix $\Pb$ can be written as $\Pb=(\be_{j_1},...,\be_{j_{s/2}}, \be_{j_{s/2+1}},$  $...,\be_{j_p})$, where the indexes $j_1,...,j_{s/2}$ belong to $[j]$ and $j_{s/2+1},...,j_p$ are not in $[j]$. Combining the above argument, the estimand $\btheta^j$ under hypothesis $H_j$ is given by
\begin{align}
\btheta^j&=\Pb\Bb^{-1}\Pb^{T}\EE_j[Xf(X)]=\Pb\Bb^{-1}\begin{pmatrix}
\rho\Phi\bm{1}_{s/2}\\
\btheta^*_{-[j]}
\end{pmatrix}=\Pb
\begin{pmatrix}
\rho\Phi\Ab^{-1}{\mathbf{1}_s}\\
\btheta^*_{-[j]}
\end{pmatrix}\nonumber\\
&=\begin{cases}
    \frac{\rho\Phi}{1+(\frac{s}{2}-1)\rho^2}, &\mbox{ if } \bw^j_\ell=1,\\
   \btheta^*_l, &\mbox{ if } \bw^j_\ell=0.
    \end{cases},
\label{eq_thm_lowerbound_2}
\end{align}
from which we know $\btheta^j$ is $s$-sparse since $\norm{\btheta^*}_0 \leq \frac{s}{2}$.

In the sequel, we will verify $\EE_j(f(X)-X^T\btheta^j)^2\leq \Phi^2$ holds. Recall that $N=(N_1,...,N_p)\sim \mathcal{N}(0,I_p)$. By (\ref{eq_thm_lowerbound_2}), we have
\begin{align*}
\EE_j(f(X)-X^T\btheta^j)^2
   &= \EE_0\Big[(f(N)-N^T\btheta^*-\frac{s\rho^2f^{\perp}(N)}{2(1+(s/2-1)\rho^2)}-\frac{\rho\Phi\sqrt{1-\rho^2}}{1+(s/2-1)\rho^2}\sum_{k\in [j]}N_k)^2\Big] \nonumber \\
  &=(1-\frac{s\rho^2}{2(1+(s/2-1)\rho^2)})^2\Phi^2+\frac{s\rho^2\Phi^2(1-\rho^2)}{2(1+(s/2-1)\rho^2)^2}\\
  &= \Phi^2\frac{(1-\rho^2)^2+s\rho^2(1-\rho^2)/2}{(1+(s/2-1)\rho^2)^2} \\
  &= \Phi^2\frac{1-(s/2-1)\rho^4+(s/2-2)\rho^2}{1+(s-2)\rho^2+(s/2-1)^2\rho^4} <\Phi^2
\end{align*}
Due to the choice of $b$, under $H_0$ we have $\EE_0(f(X)-X^T\btheta^0)^2= \Phi^2$. Therefore, we have shown that the distribution of $(X,Y)$ under the hypotheses $H_j$ for $j=0,...,|\cM'|$ belongs to the class of distributions $\mathcal{P}_{\Phi,\sigma}$.

To apply Theorem 2.7 in \cite{tsy}, we need to (1) lower bound $\|\btheta^j-\btheta^{j'}\|_q$ for $0\leq j<j'\leq |\cM'|$ and (2) upper bound the Kullback-Leibler divergence between the probability measure of the data denoted by $\mathcal{P}_j$ and $\mathcal{P}_0$ under $H_j$ and $H_0$. For (1), we have from (\ref{eq_thm_lowerbound_2}) that for $1\leq j<j'\leq |\cM'|$
\begin{equation}\label{eq_thm_lowerbound_3}
\|\btheta^j-\btheta^{j'}\|_q=\frac{\Phi\rho}{1+(s/2-1)\rho^2}\rho_H^{1/q}(\bw^j,\bw^{j'}) \geq \frac{s^{1/q}\Phi \rho}{16^{1/q}(1+(s/2-1)\rho^2)},
\end{equation}
where the last step follows from (\ref{eq_thm_lowerbound_0}). For $j=0$ and $j'\geq 1$ we have
\begin{equation}\label{eq_thm_lowerbound_4}
\|\btheta^j-\btheta^{j'}\|_q=\frac{\Phi\rho s^{1/q}}{2^{1/q}(1+(s/2-1)\rho^2)}.
\end{equation}
To quantify the Kullback-Leibler divergence, recall that the data in matrix form can be written as $(\bY,\tilde {\bX})$, where $\tilde {\bX}=(X_1,...,X_{n+N})^T$ and $\bY=(Y_1,...,Y_{n})^T$. With a slight change of notation, we use $X_{i\ell}$ to denote the $\ell$-th component of $X_i$ for $1\leq \ell\leq p$ and $1\leq i\leq n+N$. Under $H_j$, the data distribution can be decomposed as
\begin{align*}
p_j(\bY,\tilde {\bX})&=p(\bY|\bX)p_j(\tilde {\bX})=p(\bY|\bX)\prod_{i=1}^{n+N} p_j(X_i)\\
&=p(\bY|\bX)\prod_{i=1}^{n+N} p_j(X_{i,-\{[j],u\}})p_j(X_{i,[j]}|X_{i,u})p_j(X_{i,u}),
\end{align*}
where we note that the p.d.f. $p(\bY|\bX)$ remains the same across $j$ and $X_{i,-\{[j],u\}}$ stands for the subvector of $X_i$ by excluding the indexes in $\{[j],u\}$. From the above decomposition, the Kullback-Leibler divergence is given by
$$\begin{aligned}
\mathcal{K}(\mathcal{P}_j,\mathcal{P}_0)&=\EE_{j}\Big[\log\frac{p(\bY|\bX)p_j(\tilde {\bX})}{p(\bY|\bX)p_0(\tilde {\bX})}\Big]=(n+N)\EE_j\Big[\log \frac{p_j(X_{i,[j]}| X_{i,u})}{p_0(X_{i,[j]}| X_{i,u})}\Big].
\end{aligned}$$
Furthermore, notice that
$$\begin{aligned}
p_j(X_{i,[j]}|X_{i,u})&=\prod_{k\in [j]}p_j(X_{ik}|X_{i,u})\\
&=\prod_{k\in [j]}\Big(\frac{1}{\sqrt{2\pi(1-\rho^2)}}\exp\Big\{-\frac{(X_{ik}-\frac{\rho}{\Phi}f^{\perp}(X_u))^2}{2(1-\rho^2)}\Big\}\Big),
\end{aligned}$$
$$
p_0(X_{i,[j]}|X_{i,u})=\prod_{k\in [j]}p_0(X_{ik}|X_{i,u})=\prod_{k\in [j]}\Big(\frac{1}{\sqrt{2\pi}}\exp\Big\{-\frac{X_{ik}^2}{2}\Big\}\Big).$$
Hence, with some index $k\in [j]$ we obtain
\begin{align}\label{eq_thm_lowerbound_5}
\mathcal{K}(\mathcal{P}_j,\mathcal{P}_0)&=\frac{(n+N)s}{2}\Big(\EE_j[-\frac{(X_{ik}-\frac{\rho}{\Phi}f^{\perp}(X_u)))^2}{(1-\rho^2)}]-\log(1-\rho^2)+\EE_j[X_{ik}^2]\Big)\nonumber\\
&=\frac{(n+N)s}{2}\Big(1-\log(1-\rho ^2)\\
&-\frac{1}{(1-\rho^2)}\EE_j\big[X_{ik}^2+\frac{\rho^2}{\Phi^2}((f^{\perp}(X_u))^2-2\frac{\rho}{\Phi} X_{ik}f^{\perp}(X_u)\big]\Big)\nonumber\\
&= \frac{(n+N)s}{2} \big(1-\log(1-\rho ^2)-\frac{1+\rho^2-2\rho^2}{1-\rho^2}\big) \nonumber\\
&=\frac{(n+N)s}{2}\log\Big(1+\frac{\rho^2}{1-\rho^2}\Big)\nonumber \leq \frac{s(n+N)\rho^2}{2(1-\rho^2)}.
\end{align}
We set
\begin{equation}\label{eq_thm_lowerbound_rho}
\rho=\frac{1}{4}\sqrt{\frac{c_1'\log(p/s)}{n+N}},
\end{equation}
where $c_1'$ is specified in (\ref{eq_thm_lowerbound_0}). Given the condition $\frac{(s/2-1)c_1'\log(p/s)}{(n+N)}\leq 1$ and $s\geq 3$, we have $1-\rho^2\geq \frac{1}{2}$. Then from (\ref{eq_thm_lowerbound_5}) the Kullback-Leibler divergence can be bounded as follows $$\mathcal{K}(\mathcal{P}_j,\mathcal{P}_0)\leq\frac{s(n+N)\rho^2}{2(1-\rho^2)}\leq \frac{ c_1's \log (p/s)}{16}\leq\frac{1}{16}\log|\mathcal{M'}|,
$$
where the last inequality follows from (\ref{eq_thm_lowerbound_0}). Finally, with the choice of $\rho$ in (\ref{eq_thm_lowerbound_rho}), we obtain from (\ref{eq_thm_lowerbound_3}) and (\ref{eq_thm_lowerbound_4}) that
$$
\|\btheta^j-\btheta^{j'}\|_q\geq Cs^{1/q}\Phi\sqrt{\frac{\log(p/s)}{n+N}}.
$$
where we use the inequality $1+(s/2-1)\rho^2\leq 17/16$ and $C$ is a constant. We complete the proof by applying Theorem 2.7 in \cite{tsy}.
}

\subsubsection{Proof of Theorem \ref{mainthm}}
\begin{proof} Here, we show a general version with surrogate function $h(X)$. First, we can rewrite that
\begin{align*}
  \hat \bxi_j&= \sum_{i\in D^*_j}\frac{X_iY_i- \hat h^{-j}(X_i)X_i}{n_j}  -\sum_{i\in D_j}\frac{\hat h^{-j}(X_i)X_i}{n_j+N_j}\\
   &=\sum_{i\in D^*_j}\frac{X_iY_i- f(X_i)X_i}{n_j}  -\sum_{i\in D_j}\frac{f(X_i)X_i}{n_j+N_j}+ \sum_{i\in D^*_j}\frac{(f(X_i)-h(X_i)) X_i}{n_j}\\
   &-\sum_{i\in D_j}\frac{(f(X_i)-h(X_i)) X_i}{n_j+N_j}+\sum_{i\in D^*_j}\frac{(h(X_i)-\hat h^{-j}(X_i))X_i}{n_j} -\sum_{i\in D_j}\frac{(h(X_i)-\hat h^{-j}(X_i))X_i}{n_j+N_j}.
\end{align*}
Denote
\begin{equation}\label{repI}
\Rmnum{1}_1=\sum_{i\in D^*_j}\frac{(h(X_i)-\hat h^{-j}(X_i))X_i}{n_j} -\sum_{i\in D_j}\frac{(h(X_i)-\hat h^{-j}(X_i))X_i}{n_j+N_j}.
\end{equation}

\begin{equation}\label{repII}
\Rmnum{1}_2=\sum_{i\in D^*_j}\frac{(f(X_i)-h(X_i))X_i}{n_j} -\sum_{i\in D_j}\frac{(f(X_i)-h(X_i))X_i}{n_j+N_j}.
\end{equation}

Next, we aim to show that $\norm{\Rmnum{1}_1}_\infty=O_p(K_1b_n\sqrt{\frac{\log p}{n}})$. To this end, we further decompose $\Rmnum{1}$ as
\begin{align}
  \norm{\Rmnum{1}_1}_\infty  &\leq\Big\|\frac{1}{n_j}\sum_{i\in D^*_j}\{( h(X_i)-\hat h^{-j}(X_i))X_i\}-\EE_{D^*_{-j}}\big[(h(X)-\hat h^{-j}(X))X\big]\Big\|_\infty \nonumber\\
  &~~+\Big\|\frac{1}{n_j+N_j}\sum_{i\in D_j}\{(h(X_i)-\hat h^{-j}(X_i))X_i\}-\EE_{D^*_{-j}}\big[(h(X)-\hat h^{-j}(X))X\big]\Big\|_\infty,\label{eq_pf_mainthm0}
\end{align}
where $\EE_{D^*_{-j}}$ denotes the conditional expectation given the data in $D^*_{-j}=D^*\backslash D^*_j$. Let us denote $g_k(X)=(h(X)-\hat h^{-j}(X))X_k$ and $\gamma_k(X)=g_k(X)-\EE_{D^*_{-j}}[g_k(X)]$. From (\ref{eq_pf_mainthm0}), it suffices to upper bound $\max_{1\leq k\leq p}\frac{1}{n_j}\sum_{i\in D^*_j} \gamma_k(X_i)$ and $\max_{1\leq k\leq p}\frac{1}{n_j+N_j}\sum_{i\in D_j} \gamma_k(X_i)$, respectively.

We know
\begin{equation}\label{eq_pf_mainthm1}
\EE_{D^*_{-j}}\Big[\max_{1\leq k\leq p}\sum_{i\in D^*_j} g_k^2(X_i)\Big]\leq n_jK_1^2\norm{\hat h^{-j}-h}_2^2,
\end{equation}
which follows from $\|X\|_\infty\leq K_1$ in Assumption \ref{assumption_est}.  Therefore, with the application of lemma \ref{repeated_lemma} by choosing $m=2$,
we can show
$$\EE_{D^*_{-j}}\Big[\max_{1\leq k\leq p}\Big|\frac{1}{n_j}\sum_{i\in D^*_j}\gamma_k(X_i)\Big|^2\Big]\leq  K_1^2\frac{8\log(2p)}{n_j}\norm{\hat h^{-j}-h}_2^2.
$$
Furthermore, the Markov inequality implies for any $c>0$
\begin{align}
&\PP\Big(\max_{1\leq k\leq p}|\frac{1}{n_j}\sum_{i\in D^*_j}\gamma_k(X_i)|\geq cK_1b_n\sqrt{\frac{\log 2p}{n_j}}\Big|    D^*_{-j}\Big)\nonumber\\
&\leq \Big(\frac{\EE_{D^*_{-j}}[\max_{1\leq k\leq p}|\frac{1}{n_j}\sum_{i\in D^*_j}\gamma_k(X_i)|^2]}{c^2K_1^2b_n^2}{\frac{n_j}{\log (2p)}}\Big)\wedge 1\nonumber \\
&\leq \Big(\frac{8\norm{\hat h^{-j}-h}_2^2}{c^2b_n^2}\Big)\wedge 1.\label{eq_pf_mainthm2}
\end{align}
For any $\epsilon>0$, let $c'$ be a sufficiently large constant such that the event $\cE=\{\norm{\hat h^{-j}-h}_2^2\leq c'b_n^2\}$ holds with probability at least $1-\epsilon$. From (\ref{eq_pf_mainthm2}), we know that
\begin{align*}
&\PP\Big(\max_{1\leq k\leq p}|\frac{1}{n_j}\sum_{i\in D^*_j}\gamma_k(X_i)|\geq cK_1b_n\sqrt{\frac{\log 2p}{n_j}}\Big)\\
&=\EE\Big[\PP\Big(\max_{1\leq k\leq p}|\frac{1}{n_j}\sum_{i\in D^*_j}\gamma_k(X_i)|\geq cK_1b_n\sqrt{\frac{\log 2p}{n_j}}\Big|    D^*_{-j}\Big)\Big]\\
&\leq \EE\Big[(\frac{8\norm{\hat h^{-j}-h}_2^2}{c^2b_n^2}\wedge 1)I(\cE)\Big] +\EE\Big[(\frac{8\norm{\hat h^{-j}-h}_2^2}{c^2b_n^2}\wedge 1) I(\cE^c)\Big]\\
&\leq \frac{8c'}{c^2}+\PP(\cE^c)\leq 2\epsilon.
\end{align*}
where the last step holds by taking $c^2=8c'/\epsilon$ and the definition $\PP(\cE^c)\leq\epsilon$. This implies
$$
\max_{1\leq k\leq p}\Big|\frac{1}{n_j}\sum_{i\in D^*_j} \gamma_k(X_i)\Big|=O_p(K_1b_n\sqrt{\frac{\log p}{n_j}}).
$$
Following the same argument, the following probability bound holds
$$
\max_{1\leq k\leq p}\Big|\frac{1}{n_j+N_j}\sum_{i\in D_j} \gamma_k(X_i)\Big|=O_p(K_1b_n\sqrt{\frac{\log p}{n_j+N_j}}).
$$
What is more, if we denote $\norm{f-h}_2^2=\EE[(f(X)-h(X))^2]$, we can also derive
$$\norm{\Rmnum{1}_2}_\infty=O_p(K_1\norm{f-h}_2\sqrt{\frac{\log p}{n_j}}).
$$
The rest of the proof follows the same line as in the proof of Theorem 7.1 in \cite{bickel2009}. Recall that we assume for $j=\{1,2\}$, $n_j=n/2$ and $N_j=N/2$. We can show that
 \begin{align} \label{a9}
 &~~~~\Big\|\frac{\sum_{i =1 }^{n+N}X_iX_i^T}{(n+N)}\btheta^*-\frac{\hat\bxi_1+\hat \bxi_2}{2}\Big\|_\infty  \nonumber\\
 &=\Big\|\frac{\sum_{i =1 }^{n+N}X_i(X_i^T\btheta^*- f(X_i))}{(n+N)}+\frac{\sum_{i=1}^{n}X_i^T(Y_i- f(X_i))}{n}\Big\|_\infty +O_p(K_1b_n\sqrt{\frac{\log p}{n}})\nonumber\\
 &~~~+O_p(K_1\norm{f-h}_2\sqrt{\frac{\log p}{n_j}})\nonumber \\
& \leq\Big\|\frac{\sum_{i =1 }^{n+N}X_i(X_i^T\btheta^*- f(X_i))}{(n+N)}\Big\|_\infty+\Big\|\frac{\sum_{i=1}^{n}X_i^T(Y_i- f(X_i))}{n}\Big\|_\infty +O_p(K_1b_n\sqrt{\frac{\log p}{n}})\nonumber\\
&~~~+O_p(K_1\norm{f-h}_2\sqrt{\frac{\log p}{n_j}})\nonumber \\
&= O_p(K_1\Phi\sqrt{\frac{\log p}{n+N}})+O_p(K_1\sigma\sqrt{\frac{\log p}{n}})+O_p(K_1b_n\sqrt{\frac{\log p}{n}})+O_p(K_1\norm{f-h}_2\sqrt{\frac{\log p}{n_j}}).
 \end{align}
where the last probability bound holds by the same argument in the proof of example 14.3 in \cite{shd}.

Then, if we set $\lambda_{SD}= C'K_1(\Phi\sqrt{\frac{\log p}{n+N}}+(\sigma+b_n+\norm{f-h}_2)\sqrt{\frac{\log p}{n}})$ for sufficiently large $C'$, (\ref{a9}) implies
\begin{equation}
 \label{dantzig1}
     \lambda_{SD}\geq  \Big\|\frac{\sum_{i =1 }^{n+N}X_iX_i^T}{(n+N)}\btheta^*-\frac{\hat\bxi_1+\hat \bxi_2}{2}\Big\|_\infty
 \end{equation}
holds with probability tending to 1. Let $\bdelta=\hat \btheta_{SD}-\btheta^*$. By the construction of Dantzig estimator, when (\ref{dantzig1}) holds, we have $\norm{\bdelta_{T^c}}_1\leq \norm{\bdelta_T}_1$ where $T$ denotes the support of $\btheta^*$ and
$$\begin{aligned}
\frac{1}{n+N}\norm{\tilde{\bX^T}\tilde {\bX}\bdelta}_\infty\leq \norm{\frac{\tilde{\bX^T}\tilde {\bX}\hat \btheta_{SD}}{n+N}-\frac{\hat\bxi_1+\hat \bxi_2}{2}}_\infty+\norm{\frac{\tilde{\bX^T}\tilde {\bX}\btheta^*}{n+N}-\frac{\hat\bxi_1+\hat \bxi_2}{2}}_\infty\leq 2\lambda_{SD}.
\end{aligned}$$
Therefore,
$$\begin{aligned}
\frac{1}{n+N}\norm{\tilde{\bX}\bdelta}^2_2&=\frac{1}{n+N}\bdelta^T\tilde{\bX^T}\tilde{\bX}\bdelta\\
&\leq \frac{1}{n+N}\norm{\bdelta^T\tilde{\bX^T}\tilde{\bX}}_\infty\norm{\bdelta}_1\\
&\leq 2\lambda_{SD}\times 2\norm{\bdelta_T}_1\leq 4\lambda_{SD}\sqrt{s}\norm{\bdelta_T}_2.
\end{aligned}$$
With smallest eigenvalue condition (A1) and Lemma \ref{RE}, we know on the event of $\mathcal{B}_{c_1}(n+N,s,1)$, $\frac{1}{n+N}\norm{\tilde{\bX}\bdelta}_2^2\geq (1-c_1)^2C_{\min} \norm{\bdelta_T}_2^2$. Therefore,
$$\norm{\bdelta_T}_2\leq \frac{4\lambda_{SD}\sqrt{s}}{(1-c_1)^2C_{\min}}.
$$
Above all, we know $\norm{\bdelta}_1\leq 2\norm{\bdelta_T}_1\leq 2\sqrt{s}\norm{\bdelta_T}_2\leq \frac{8\lambda_{SD} s}{(1-c_1)^2C_{\min}}$. As a byproduct, we can show that $\norm{\bdelta}_2\lesssim \lambda_D\sqrt{s}$. This completes the proof.
\end{proof}

\subsubsection{Proof of Theorem \ref{twostep}}

Before the proof starts, we need an additional notation for the largest restricted eigenvalue. Denote
$$\phi_{\max} (L) = \max_{1\leq \norm{v}_0\leq L} \frac{v^T\hat \bSigma_{n+N} v}{\norm{v}_2^2}.$$
Then we know $\phi_{\max}(\cdot)$ is an increasing function. We further use the same notation to denote the event for the upper bound parallel to the lower bound in definition (\ref{lowerRE})
\begin{equation}
     \mathcal{B}^U_\delta(n,s_0,L)=\{\bX\in \RR^{n\times p}:\frac{\norm{\bX \bv}_2}{\sqrt{n}\norm{\bv}_2}\leq (1+\delta)\sqrt{C_{\max}} ,\forall \bv\in \mathcal{C}(s_0,L)~\textrm{s.t.}~\bv\neq 0\},
 \end{equation}
and under same condition in Lemma \ref{RE},
\begin{equation}
\PP(\mathcal{B}^U_\delta(n,s_0,L))\geq 1-e^{-\delta^2n}.
\end{equation}

Lemma 3.5 in \cite{debiased} shows that $|\hat{T_1}|\lesssim s$ with high probability under the assumption that $s(\log p)^2=o(n)$. With those notations above, we can show the following lemma ensuring the sparsity of $\hat T_2$ in a similar spirit.
\begin{lem} \label{t2sparse}
Suppose Assumption \ref{assumption_est} holds, $\Lambda_{\max}(\bSigma)\leq C_{\max}<\infty$ and $s(\log p)^2=o(n+N)$. On the event of $\mathcal{B}_\delta(n,s,3)\cap \mathcal{B}^U_\delta(n,s,3)\cap \big\{\lambda_{SL}\geq  2\norm{\bSigma_{n+N}\btheta^*-\hat{\bxi} }_\infty\big\}$, we can show $$|\hat T_2|\leq \frac{64C_{\max}s}{(1-c_1)^2 C_{\min}}.$$
\end{lem}

\begin{proof}
Recall the estimator $\hat \btheta_{SL}$ is defined as
\begin{equation}\label{lassotype}
    \hat \btheta_{SL}=\argmin_{\btheta \in \RR^p} \sum_{j=1}^2\frac{\sum_{i\in D_j}\big(\hat h^{-j}(X_i)-X_i^T\btheta\big)^2}{n+N} -\frac{2\sum_{i\in D_j^*}\big( Y_i-\hat h^{-j}(X_i)\big)X_i^T\btheta}{n}+2\lambda_{SL} \norm{\btheta}_1.
\end{equation}
By KKT condition, we have
\begin{align*}
    \sum_{j=1}^2\frac{\sum_{i\in D_j}\big(\hat h^{-j}(X_i)-X_i^T\hat\btheta_{SL}\big)X_i}{n+N}
     +\frac{\sum_{i\in D_j^*}\big(Y_i-\hat h^{-j}(X_i)\big)X_i}{n}=\lambda_{SL} v(\hat\btheta_{SL}), ~~v(\btheta)\in \partial \norm{\btheta}_1.
\end{align*}
Hence we know
\begin{align*}
     &\frac{\sum_{i=1}^{n+N}\big( X_i^T\btheta^*-X_i^T\hat\btheta_{SL}\big)X_i}{n+N}
     =\lambda_{SL} v(\hat\btheta_{SL}^{(j)})\\
     &~~~~~~~~~~~~-\sum_{j=1}^2\Big(\frac{\sum_{i\in D_j^*}\big(Y_i-\hat h^{-j}(X_i)\big)X_i}{n}+  \frac{\sum_{i\in D_j}\big(\hat h^{-j}(X_i)-\hat X_i^T\btheta^*\big)X_i}{n+N}\Big),
\end{align*}
and by previous reasoning in Theorem \ref{mainthm}, we know
$$\norm{\sum_{j=1}^2\Big(\frac{\sum_{i\in D_j^*}\big(Y_i-\hat h^{-j}(X_i)\big)X_i}{n}+  \frac{\sum_{i\in D_j}\big(\hat h^{-j}(X_i)-\hat X_i^T\btheta^*\big)X_i}{n+N}\Big)}_\infty\leq \frac{\lambda_{SL}}{2}.$$
Therefore, for $k\in \hat T_2$, $ \big|\frac{\sum_{i\in D_j}\big( X_i^T\btheta^*-X_i^T\hat\btheta_{SL}\big)X_i^k}{n+N}\big|
     \geq \lambda_{SL}/2$. We square both sides of the inequality and sum over $k \in \hat T_2$. With $\Delta=\tilde{\bX}(\btheta^*-\hat\btheta_{SL})$, we attain that
\begin{equation}\label{sparsity}
    \frac{\lambda_{SL}^2}{4}|\hat T_2|\leq    \frac{\sum_{k \in \hat T_2}(\Delta^T\tilde{\bX_{\cdot k}})^2}{n+N}\leq \norm{(\hat\bSigma_{n+N})_{\hat T_2}}_2^2\norm{\Delta}_2^2.
\end{equation}
First, by a standard argument for lasso estimator, we can show the same result as for the dantzig selector in Theorem \ref{mainthm} and we know $\norm{\Delta}_2^2\leq 16s\lambda_{SL}^2/\big((1-c_1)^2 C_{\min} \big)$ on the event of $\mathcal{B}_\delta(n,s,3)\cap \big\{\lambda_{SL}\geq  2\norm{\hat\bSigma_{n+N}\btheta^*-\hat{\bxi} }_\infty\big\}$ .

Second, employing Lemma 3.5 in \cite{debiased}, we know
\begin{equation}
    \PP(\phi_{\max}(k)\geq C_{\max}+C\sqrt{\frac{k}{n+N}}+\frac{t}{\sqrt{n+N}})\leq 2\exp{(-ct^2+k\log p+k)},
\end{equation}
for $t\geq 0$, where $C,c$ depend only on $C_{\max}$.
Hence, $\norm{(\hat\bSigma_{n+N})_{\hat T_2}}_2\leq\phi_{\max}(|\hat T_2|)\leq \phi_{\max}(n+N) \leq c_1\sqrt{\frac{(n+N)\log p}{n+N}}$ holds with large probability, and by inequality (\ref{sparsity}), we know $|\hat T_2|\lesssim s\sqrt{\log p}$. Finally, we can refine the bound, as $\phi_{\max}(|\hat T_2|)\leq \phi_{\max}(C's\sqrt{\log p})\leq C_{\max}$ if $s(\log p)^2=o(n+N)$. From (\ref{sparsity}), we derive that
$$|\hat T_2|\leq \frac{64C_{\max}s}{(1-c_1)^2 C_{\min}}.$$
\end{proof}

Now, we are ready to prove Theorem \ref{twostep}.

\begin{proof}

 We take $\bomega=0$ in the basic inequality,
\begin{align*}
    \frac{1}{2n}\sum_{i=1}^n(Y_i-X_i^T(\hat \btheta_{SL}+ \hat\bomega))^2+\lambda_\omega\norm{\hat\bomega}_1\leq  \frac{1}{2n}\sum_{i=1}^n(Y_i-X_i^T\hat \btheta_{SL})^2.
\end{align*}
Then we have
\begin{align*}
    \frac{1}{2n}\sum_{i=1}^n(X_i^T\hat\bomega)^2&+\lambda_{\omega}\norm{\hat\bomega}_1\leq  \frac{1}{n}|\sum_{i=1}^n(Y_i-X_i^T\hat \btheta_{SL})X_i^T\hat{\bomega}|\\
    &\leq \frac{1}{n}|\sum_{i=1}^n(Y_i-X_i^T \btheta^*)X_i^T\hat{\bomega}|+\frac{1}{n}|\sum_{i=1}^nX_i^T(\btheta^*-\hat \btheta_{SL})X_i^T\hat{\bomega}|\\
    &\leq \norm{\frac{1}{n}\sum_{i=1}^n(Y_i-X_i^T \btheta^*)X_i^T}_\infty\norm{\hat\bomega}_1+\frac{4\sum_{i=1}^n(X_i^T(\btheta^*-\hat \btheta_{SL}))^2}{n}\\
    &~~~+\frac{1}{4n}\sum_{i=1}^n(X_i^T\hat\bomega)^2.
\end{align*}
Denote $\lambda_0=\norm{\frac{1}{n}\sum_{i=1}^n(Y_i-X_i^T \btheta^*)X_i^T}_\infty$. Choosing $\lambda_{\omega}\geq C_\omega K_1(\Phi+\sigma)\sqrt{\frac{\log p}{n}}\geq 2\lambda_0$, we then arrive that
\begin{align*}
 \frac{1}{4n}\sum_{i=1}^n(X_i^T\hat\bomega)^2+\frac{\lambda_\omega}{2}\norm{\hat \bomega}_1\leq \frac{4}{n}\sum_{i=1}^n(X_i^T(\btheta^*-\hat \btheta_{SL}))^2.
\end{align*}
With the given sparsity of $\supp(\hat\bomega)$, we can proceed to show the rate of $\hat\btheta_{SL}$ hold for $\hat \btheta_R$ by
\begin{align*}
    (1-\delta)C_{\min}\norm{\hat \bomega}_2^2\leq 16(1+\delta)C_{\max}\norm{\btheta^*-\hat\btheta_{SL}}_2^2.
\end{align*}
That is to say $\norm{\hat\btheta_R-\btheta^*}_2\leq \norm{\btheta^*-\hat \btheta_{SL}}_2+\norm{\hat \bomega}_2\leq (\frac{2C_{\max}}{C_{\min}}+1) \norm{\btheta^*-\hat \btheta_{SL}}_2$.

From the basic inequality, taking $\omega=\hat \btheta_L-\hat\btheta_{SL}$,
\begin{align*}
    \frac{1}{2n}\sum_{i=1}^n(Y_i-X_i^T(\hat \btheta_{SL}+ \hat\bomega))^2+\lambda_\omega\norm{\hat\bomega}_1\leq  \frac{1}{2n}\sum_{i=1}^n(Y_i-X_i^T\hat \btheta_L)^2+\lambda_\omega\norm{\hat \btheta_L-\hat\btheta_{SL}}_1,
\end{align*}
we can see
\begin{align*}
    \frac{1}{2n}\sum_{i=1}^n(Y_i-X_i^T\hat \btheta_R)^2\leq  \frac{1}{2n}\sum_{i=1}^n(Y_i-X_i^T\hat \btheta_L)^2+\lambda_\omega(\norm{\hat \btheta_L-\hat\btheta_{SL}}_1-\norm{\hat\bomega}_1).
\end{align*}
Therefore, we have
\begin{align*}
    \frac{1}{2n}\sum_{i=1}^n (X_i^T(\hat \btheta_L-\hat\btheta_R))^2\leq \frac{1}{n}|\sum_{i=1}^n(Y_i-X_i^T\hat \btheta_L)X_i^T(\hat \btheta_L-\hat\btheta_R)|+ \lambda_\omega\norm{\hat \btheta_L-\hat\btheta_R}_1.
\end{align*}
We know with the same choice of $\lambda_\omega\geq CK_1(\Phi+\sigma)\sqrt{\frac{\log p}{n}}$ with large enough constant $C$, $$\lambda_\omega \geq\norm{\frac{\sum_{i=1}^n(Y_i-X_i^T\hat \btheta_L)X_i}{n}}_\infty=O_p(K_1(\Phi+\sigma)\sqrt{\frac{\log p}{n}}),$$
so we attain that
$$ \frac{1}{2n}\sum_{i=1}^n (X_i^T(\hat \btheta_L-\hat\btheta_R))^2\leq 2\lambda_\omega\norm{\hat \btheta_L-\hat\btheta_R}_1.$$
The support set of $\hat\btheta_L-\hat\btheta_R$ is contained in $\hat T_1\cup \hat T_2$, so we have $$\norm{\hat \btheta_L-\hat\btheta_R}_1\lesssim \sqrt{s}\norm{\hat \btheta_L-\hat\btheta_R}_2.$$
On the event of $\mathcal{B}_{c_1}(n,|\hat T_1\cup \hat T_2|,1)$,
$$\frac{1}{n}\sum_{i=1}^n (X_i^T(\hat \btheta_L-\hat\btheta_R))^2\geq (1-c_1)^2C_{\min} \norm{\hat \btheta_L-\hat\btheta_R}_2^2.$$
Hence, $\norm{\hat \btheta_L-\hat\btheta_R}_2\leq \lambda_\omega \sqrt{s}/((1-c_1)^2C_{\min}).$ What is more, $\norm{\hat \btheta_R-\btheta^*}_2\leq \norm{\hat \btheta_R-\hat \btheta_L}_2+\norm{\hat \btheta_L-\btheta^*}_2\lesssim\lambda_\omega\sqrt{s}$.


\end{proof}

\subsubsection{Proof of Proposition \ref{twostep_h}}
The proof is in a similar spirit as Theorem \ref{twostep}. We only provide a sketch.

\begin{proof}
Recall that the objective function can be rewritten as  $$(\hat \btheta_{h_1}+\bomega)^T\hat \bSigma_{n+N}(\hat \btheta_{h_1}+\bomega)-(\hat \bxi_1+\hat \bxi_2)^T(\hat \btheta_{h_1}+\bomega)+2\lambda_H \norm{\bomega}_1.$$
where $\hat \bxi$ is defined with $h_2$. First, we can take $\bomega=0$ in the basic inequality, so
\begin{align*}
 (\hat \btheta_{h_1}+\hat\bomega)^T\hat \bSigma_{n+N}(\hat \btheta_{h_1}+\hat\bomega)-(\hat \bxi_1+\hat \bxi_2)^T\hat \btheta_{R}+2\lambda_H \norm{\hat\bomega}_1\leq \hat \btheta_{h_1}^T\hat \bSigma_{n+N}\hat \btheta_{h_1}-(\hat \bxi_1+\hat \bxi_2)^T \hat \btheta_{h_1}.
\end{align*}
Simplify it and we have
 \begin{align*}
  \hat \bomega^T\hat \bSigma_{n+N}\hat\bomega +2&\lambda_H\norm{\hat\bomega}_1\leq 2|\hat\bomega^T(\hat \bSigma_{n+N}\hat \btheta_{h_1}-\frac{(\hat \bxi_1+\hat \bxi_2)}{2})|\\
  &\leq 2\norm{(\hat \bSigma_{n+N} \btheta^*-\frac{(\hat \bxi_1+\hat \bxi_2)}{2})}_\infty\norm{\hat \bomega}_1+2|\hat \bomega\hat \bSigma_{n+N}(\btheta^*-\hat \btheta_{h_1})|\\
  &\leq \lambda_H\norm{\hat \bomega}_1+\frac{\hat \bomega^T\hat \bSigma_{n+N}\hat\bomega}{2}+2(\btheta^*-\hat \btheta_{h_1})^T\hat \bSigma_{n+N}(\btheta^*-\hat \btheta_{h_1})
 \end{align*}
 Hence we have $$ \frac{1}{2}\hat \bomega^T\hat \bSigma_{n+N}\hat\bomega +\lambda_H\norm{\hat\bomega}_1\leq 2(\btheta^*-\hat \btheta_{h_1})^T\hat \bSigma_{n+N}(\btheta^*-\hat \btheta_{h_1}) $$
 As what we did previously, we choose $\lambda_H\geq 2\norm{\hat \bSigma_{n+N} \btheta^*-\frac{(\hat \bxi_1+\hat \bxi_2)}{2}}_\infty$ as the one computing $\hat \btheta_{h_2}$.
Therefore, $\norm{\hat \bomega}_2 \leq \sqrt{C_{\max}/C_{\min}}\norm{\btheta^*-\hat \btheta_{h_1}}_2$ and  $\norm{\hat \btheta_R-\btheta^*}_2\leq \big(\sqrt{C_{\max}/C_{\min}}+1\big)\norm{\btheta^*-\hat \btheta_{h_1}}_2$.\\

 On the other hand, we choose $\bomega=\hat \btheta_{h_2}-\hat \btheta_{h_1}$. Then
 \begin{align*}
 (\hat \btheta_{h_1}+\hat\bomega)^T\hat \bSigma_{n+N}(\hat \btheta_{h_1}+\hat\bomega)&-(\hat \bxi_1+\hat \bxi_2)^T\hat \btheta_{R}+2\lambda_H \norm{\hat\bomega}_1\leq \\
 &\hat \btheta_{h_2}^T\hat \bSigma_{n+N}\hat \btheta_{h_2}-(\hat \bxi_1+\hat \bxi_2)^T \hat \btheta_{h_2}+2\lambda_H \norm{\hat  \btheta_{h_2}-\hat  \btheta_{h_1}}_1.
\end{align*}
It simplifies to
\begin{align*}
 (\hat \btheta_{AH}-\hat \btheta_{h_2})^T\hat \bSigma_{n+N}(\hat \btheta_{AH}&-\hat \btheta_{h_2})+2\lambda_H \norm{\hat\bomega}_1\leq\\
&2|(\hat \btheta_{AH}-\hat \btheta_{h_2})^T(\hat \bSigma_{n+N}\hat \btheta_{h_2}-\frac{(\hat \bxi_1+\hat \bxi_2)}{2})|+ 2\lambda_H \norm{\hat  \btheta_{h_2}-\hat  \btheta_{h_1}}_1
\end{align*}
Therefore, choose $\lambda_H\geq 2\norm{\hat \bSigma_{n+N}\hat \btheta_{h_2}-\frac{(\hat \bxi_1+\hat \bxi_2)}{2}}_\infty$, and we have
\begin{equation}
    (\hat \btheta_{AH}-\hat \btheta_{h_2})^T\hat \bSigma_{n+N}(\hat \btheta_{AH}-\hat \btheta_{h_2}) \leq 3\lambda_H\norm{\hat \btheta_{AH}-\hat \btheta_{h_2}}_1
\end{equation}
With the sparsity on $\hat H_1$ and $\hat H_2$, we have $\norm{\hat \btheta_{AH}-\hat \btheta_{h_2}}_2\leq 3\lambda_{h_2}\sqrt{s}/\big((1-c_1)^2C_{\min}\big)$
and $\norm{\hat \btheta_{AH}-\btheta^*}_2\leq4 \lambda_{h_2}\sqrt{s}/\big((1-c_1)^2C_{\min}\big).$
\end{proof}

\subsection{Supplementary Technical Results}\label{app_sup}

\subsubsection{$\EE(f(X)-X^T\theta^*)^2\asymp s$}\label{app_phi}

Assume that the true conditional mean function $f(X)$ has an additive form $f(X)=\sum_{k\in S}f_k(X_k)$ with $\EE(f_k(X_k))=0$, where $X_k$ is the $k$-th component of $X$ and $S$ is a subset of $\{1,...,p\}$ with $|S|=s$. We further assume that all the covariates $X_1,...,X_p$ are mutually independent. We have
\begin{align*}
\EE(f(X)-X^T\btheta^*)^2&=\EE\Big\{\sum_{k\in S}(f_k(X_k)-X_k\theta_k^*)-\sum_{k\notin S}X_k\theta_k^*\Big\}^2\\
&=\EE\Big\{\sum_{k\in S}(f_k(X_k)-X_k\theta_k^*)\Big\}^2+\EE\Big\{\sum_{k\notin S}X_k\theta_k^*\Big\}^2\\
&=\sum_{k\in S}\EE(f_k(X_k)-X_k\theta^*_k)^2+\sum_{k\notin S} \Sigma_{kk}(\theta^*_k)^2,
\end{align*}
where we use the fact that $\EE(f_k(X_k))=0$ and $\EE(X_k)=0$. By the definition of $\btheta^*$, we know that for $k\notin S$, $\theta^*_k=0$. For $k\in S$, $\theta^*_k=\argmin_{\theta} \EE(f_k(X_k)-X_k\theta)^2=\EE(f_k(X_k)X_k)/\Sigma_{kk}$ and $$\EE(f_k(X_k)-X_k\theta^*_k)^2=\EE(f_k^2(X_k))-[\EE(f_k(X_k)X_k)]^2/\Sigma_{kk}$$ which is  a constant. For example, if $X_k\sim N(0,1)$ and $f_k(X_k)=X_k^3$, then $\EE(f_k(X_k)-X_k\theta^*_k)^2=6$. Thus, we have $\EE(f(X)-X^T\btheta^*)^2=6s$.

\subsubsection{Rate of $\hat\theta_{U}$}\label{app_hathetaU}

Recall that the modified Dantzig selector $\hat\theta_{U}$ is defined as
$$
\hat\btheta_{U}=\arg\min \|\btheta\|_1,~~\textrm{s.t.}~~ \|\hat\bSigma_{n+N}\btheta-\frac{1}{n}\sum_{i=1}^{n}X_iY_i \|_\infty \leq \lambda_U.
$$
Assume that  Assumption \ref{assumption_est} holds. By choosing $\lambda_U\asymp K_1\sqrt{\frac{(\sigma^2+\Phi^2+\btheta^{*T}\bSigma\btheta^*)\log p}{n}}$, we obtain that
$$
\|\hat\btheta_U-\btheta^*\|_1=O_p\Big(sK_1(\Phi+\sigma+(\btheta^{*T}\bSigma\btheta^*)^{1/2})\sqrt{\frac{\log p}{n}}\Big).
$$
Under the conditions in our Theorem \ref{mainthm}, we have $K_1=O(1)$. This implies the bound (\ref{eq_ratehatU}).

The proof follows from the same argument as in Lemma \ref{bdant}. The only nontrivial step is to bound $\|\hat\bSigma_{n+N}\btheta-\frac{1}{n}\sum_{i=1}^{n}X_iY_i \|_\infty$. Using the triangle inequality, we have
\begin{align*}
&\|\hat\bSigma_{n+N}\btheta-\frac{1}{n}\sum_{i=1}^{n}X_iY_i \|_\infty\leq \|\frac{1}{n}\sum_{i=1}^{n}X_i(Y_i-f(X_i)) \|_\infty\\
&+ \|\frac{1}{n}\sum_{i=1}^{n}X_i(f(X_i)-X_i^T\btheta^*) \|_\infty
+ \|(\hat\bSigma_n-\bSigma)\btheta^* \|_\infty+ \|(\hat\bSigma_{n+N}-\bSigma)\btheta^* \|_\infty.
\end{align*}
We have already derived in (\ref{a9}) that,
\begin{align*}
   \|\frac{1}{n}\sum_{i=1}^{n}X_i(Y_i-f(X_i)) \|_\infty\lesssim K_1\sigma\sqrt{\frac{\log p}{n}},\\ \|\frac{1}{n}\sum_{i=1}^{n}X_i(f(X_i)-X_i^T\btheta^*) \|_\infty\lesssim K_1\Phi\sqrt{\frac{\log p}{n}}.
\end{align*}

To control $\|(\hat\bSigma_n-\bSigma)\btheta^* \|_\infty$, we note that $\|X_i\|_\infty\leq K_1$ and $\EE(X_i^T\btheta^*)^2=\btheta^{*T}\bSigma\btheta^*$. We obtain
$$
\|(\hat\bSigma_n-\bSigma)\btheta^* \|_\infty\lesssim K_1(\btheta^{*T}\bSigma\btheta^*)^{1/2}\sqrt{\frac{\log p}{n}},
$$
by the Nemirovski moment inequality in Lemma \ref{repeated_lemma} and Markov inequality. The last term $\|(\hat\bSigma_{n+N}-\bSigma)\btheta^* \|_\infty$ is dominated by $\|(\hat\bSigma_n-\bSigma)\btheta^* \|_\infty$ and can be ignored. Thus, we obtain
$$
\|\hat\bSigma_{n+N}\btheta-\frac{1}{n}\sum_{i=1}^{n}X_iY_i \|_\infty\lesssim K_1(\Phi+\sigma+(\btheta^{*T}\bSigma\btheta^*)^{1/2})\sqrt{\frac{\log p}{n}}.
$$

{
\subsubsection{Rate of $\hat\theta_{T}$}\label{app_hathetaT}

Although the method in \cite{earliest} does not apply cross-fitting procedure and they did not have a convergence rate result, to facilitate rate analysis here, we still use cross-fitting version of pseudo outcomes and let $\Tilde{\bY} = (\hat h^{-j_1} (X_1),...,\hat h^{-j_{n+N}}(X_{n+N}))$, where $j_i =1,2$ depending on which split of data $X_i$ is in.

Recall that the transductive Dantzig selector is given by
\begin{equation}\label{eq_trans}
   \hat\btheta_T=\argmin~\norm{\btheta}_1~s.t.~ \frac{1}{n+N}\norm{\tilde\bX(\Tilde{\bY}-\tilde\bX^T\btheta)}_\infty\leq \lambda_T,
\end{equation}
where $\Tilde{\bY} = (\tilde Y_1, ...,\tilde Y_{n+N})=(\hat h^{-j_1} (X_1),...,\hat h^{-j_{n+N}}(X_{n+N}))$.
Under assumption \ref{assumption_est} and conditions in Theorem \ref{mainthm}, by choosing $\lambda_T\asymp K_1\Big(\Phi\sqrt{\frac{\log p}{n+N}}+ (b_n+G_h)\Big)$, we obtain that
$$\|\hat \btheta_T -\btheta^*\|_q = O_p\Big(
K_1s^{1/q}\Big\{\Phi\sqrt{\frac{\log p}{n+N}}+ (b_n+G_h)\Big\}\Big).
$$
This rate is no better than our proposed methods when $b_n+G_h \gg (b_n+G_h+\sigma)\sqrt{\frac{\log p }{n}}$. This is true since, even in the best situation that $G_h=0$, we can expect that $b_n$ would still not be faster than $\sigma\sqrt{\frac{\log p}{n}}$.

The proof mainly follows the same argument as in Lemma
\ref{bdant}. The only nontrivial step is to bound $\frac{1}{n+N}\norm{\tilde\bX(\Tilde{\bY}-\tilde\bX^T\btheta)}_\infty$.
Using triangle inequality, we have
\begin{align*}
    \frac{1}{n+N}\norm{\tilde\bX(\Tilde{\bY}-\tilde\bX^T\btheta)}_\infty \leq  \norm{\frac{\sum_{i=1}^{n+N}X_i(\tilde Y_i-f(X_i))}{n+N}}_\infty + \norm{\frac{\sum_{i=1}^{n+N}X_i(f(X_i)-X_i^T\btheta^*))}{n+N}}_\infty
\end{align*}

We have already derived in \ref{a9} that,
\begin{align*}
   \|\frac{1}{n+N}\sum_{i=1}^{n+N}X_i(f(X_i)-X_i^T\btheta^*) \|_\infty\lesssim K_1\Phi\sqrt{\frac{\log p}{n+N}}.
\end{align*}

To deal with $\norm{\frac{\sum_{i=1}^{n+N}X_i(\tilde Y_i-f(X_i))}{n+N}}_\infty$, we adopt the other formulation of this term, which is
\begin{align*}
    \norm{&\frac{\sum_{j=1}^2\sum_{i \in D_j}X_i(\hat h^{-j}(X_i)-f(X_i))}{n+N}}_\infty \leq \frac{1}{2}\sum_{j=1}^2\norm{\frac{\sum_{i \in D_j}X_i(\hat h^{-j}(X_i)-f(X_i))}{n_j+N_j}}_\infty \\
    & \leq \frac{1}{2}\big(\sum_{j=1}^2\norm{\frac{\sum_{i \in D_j}X_i(\hat h^{-j}(X_i)-f(X_i))}{n_j+N_j}-\EE_{D^*_{-j}}\big[(\hat h^{-j}(X)-f(X))X\big]}_\infty \\
    &+ \|\EE_{D^*_{-j}}\big[(\hat h^{-j}(X)-f(X))X\big]\|_\infty\big)\\
    &=O_p\big(K_1(b_n+G_h)\sqrt{\frac{\log{p}}{n+N}}+K_1(b_n+G_h) \big) = O_p\big(K_1(b_n+G_h)\big),
\end{align*}

where the bounds are derived similarly to \ref{repI} and \ref{repII}. Thus, we obtain that
$$\frac{1}{n+N}\norm{\tilde\bX(\Tilde{\bY}-\tilde\bX^T\btheta)}_\infty \lesssim K_1\big(\Phi\sqrt{\frac{\log p}{n+N}}+ (b_n+G_h)\big).$$
}

\subsubsection{Counter Example of Weighted Estimator}\label{counter}

Since \cite{Nemirovski2000}, there is a vast literature on aggregation of estimators; see \cite{rigollet2012sparse} for a recent review. In the context of Section \ref{sec_safe}, as an alternative, we may consider the weighted estimator of the form $\hat\btheta_{\beta}=\beta \hat\btheta_{SL}+(1-\beta)\hat\btheta_{L}$ for some $\beta\in [0,1]$, which corresponds to the convex combination of two estimators $\hat\btheta_{SL}$ and $\hat\btheta_{L}$. As suggested by \cite{Nemirovski2000}, to find the optimal weighted estimator, we have to  split the sample. The first sample is used to compute the initial estimators $\hat\btheta_{SL}$ and $\hat\btheta_{L}$, while the optimal weight $\hat\beta$ is estimated from the second sample, say $\{(X_1,Y_1),...,(X_{n'},Y_{n'})\}$, by solving the following constrained least square problem
\begin{equation}
    \hat \beta=\argmin_{\beta\in[0,1]} \frac{1}{n'}\sum_{i=1}^{n'} (Y_i-X_i^T(\beta \hat\btheta_{SL}+(1-\beta)\hat\btheta_{L}))^2.
\end{equation}
The optimal weighted estimator is given by $\hat\btheta_{\hat\beta}$. However, our preliminary analysis suggests that in the semi-supervised setting, an optimal performance in the prediction does not necessarily lead to a more accurate estimation of $\btheta^*$. In other words, our Theorem \ref{twostep} does not generally hold for the weighted estimator $\hat\btheta_{\hat\beta}$. To illustrate this argument, a counter example is given below. Along this line, one future research question is that, in the context of Section \ref{sec_safe2}, whether it is possible to combine an increasing number of semi-supervised estimators via some more refined aggregation methods, such as Q-aggregation \citep{aggregation}.

We next study the property of this weighted estimator. Above is a univariate least square problem and denote \begin{equation}\label{weight}
    \hat W=\frac{\sum_{i=1}^{n'}(Y_i-X_i^T\hat\btheta_{L})X_i^T(\hat \btheta_{SL}-\hat \btheta_{L})}{\sum_{i=1}^{n'}(X_i^T(\hat \btheta_{SL}-\hat \btheta_{L}))^2}.
\end{equation}
Then $$\hat \beta=\begin{cases}
\hat W &  \textrm{if} ~ \hat W\in[0,1],\\
0 & \textrm{if} ~ \hat W<0,\\
1 & \textrm{if} ~ \hat W>1.
\end{cases}$$
With $\hat \beta$, we denote $\hat \btheta_{\hat \beta}=\hat\beta \hat\btheta_{SL}+(1-\hat\beta)\hat\btheta_{L}$.

Firstly, we consider the case where $\hat \beta=0$, and $\hat \beta=1$ follows the same argument. When $\hat \beta=0$,
we have
\begin{equation}
    \sum_{i=1}^{n'}(Y_i-X_i^T\hat \btheta_{L})^2\leq  \sum_{i=1}^{n'}(Y_i-X_i^T\hat \btheta_{SL})^2,
\end{equation}
from which we can get
\begin{align}\label{agg}
    (\hat \btheta_{L}-\btheta^*)^T\hat \bSigma_n(\hat \btheta_{L}-\btheta^*)\leq  (\hat \btheta_{ SL}-\btheta^*)^T\hat \bSigma_n(\hat \btheta_{ SL}-\btheta^*)&+\frac{2}{n'}\sum_{i=1}^{n'}((\eta_i+\epsilon_i)X_i^T(\hat\btheta_{L}-\hat\btheta_{SL})).
\end{align}
The remainder term $\frac{2}{n'}\sum_{i=1}^{n'}((\eta_i+\epsilon_i)X_i^T(\hat\btheta_{L}-\hat\btheta_{SL}))$ is $O_p(\frac{K_1(\Phi+\sigma)\norm{\hat\btheta_L-\hat\btheta_{SL}}_2}{\sqrt{n}})$. Intuitively, by triangle inequality, we can show that
$\norm{\hat\btheta_L-\hat\btheta_{SL}}_2$ is of order $K_1(\sigma+\Phi)\sqrt{\frac{s\log p}{n}}$, and therefore the remainder term is $O_p(\frac{K_1^2(\Phi+\sigma)^2\sqrt{s\log p}}{n})$. However, the inequality \ref{agg} shows that $\norm{\hat\btheta_L-\btheta^*}_2$ is not guaranteed to be of smaller order than $\norm{\hat\btheta_{SL}-\btheta^*}_2$, although the prediction from $\hat\btheta_L$ is preferred.

Indeed, in this case $\norm{\hat\btheta_L-\btheta^*}_2$ can be larger than the order of $\norm{\hat\btheta_{SL}-\btheta^*}_2$. To illustrate this argument, a counter example is given as follows.

Assume that the true conditional mean function $f(X)$ is supported on $S$, a subset of $\{1,...,p\}$ with $|S|=s$ and projection coefficients are nonzero only on $S$ with value $\theta$. We further assume that all the covariates $X_1,...,X_p$ are i.i.d.uniformly distributed on [-1,1].

Suppose we have two initial estimators $\hat \btheta_L$ and $\hat \btheta_{SL}$ whose support sets are both $S\cup S'$, where $|S'|=s$. Let $\hat \btheta_L^k=\hat \btheta_{SL}^k=\theta+\delta$ for $k\in S$, $\hat \btheta_L^k=\delta_1$, $\hat \btheta_{SL}^k=\delta_2$ for $k\in S'$ and $\delta_1,\delta_2\geq 0$. Next, we construct examples where $\hat\btheta_{L}$ achieves the minimal prediction error among all the convex weighted estimator but has arbitrarily larger order of estimation error than $\hat \btheta_{SL}$. We can compute $\hat W$ according to equation (\ref{weight}), and the denominator is $\sum_{i=1}^{n'}(\sum_{k\in S'}X_i^k)^2(\delta_1-\delta_2)^2$, where $X_i^k$ is the $k$-th coordinate of $X_i$. And the numerator is $\sum_{i=1}^{n'}(Y_i-\sum_{k\in S}X_i^k(\theta-\delta)-\sum_{k\in S'}X_i^k\delta_1)(\sum_{k\in S'}X_i^k(\delta_2-\delta_1))$. Set $\delta=\sigma\sqrt{\frac{\log p}{n'}}$, $\delta_2=\Phi\sqrt{\frac{\log p}{n'+N}}$ and $\delta_1=\rho\sqrt{\frac{\log p}{n'}}$, where $N\gg n'$ and $\rho\ll \Phi$ but $\delta_1\gg\delta_2$, so that $\norm{\hat\btheta_L-\btheta^*}_2\gg \norm{\hat\btheta_{SL}-\btheta^*}_2^2$.

Denote $$T_1=\frac{1}{n'}\sum_{i=1}^{n'}(\sum_{k\in S'}X_i^k)^2(\delta_1-\delta_2)^2,$$
$$T_2=\frac{1}{n'}\sum_{i=1}^{n'}(Y_i-\sum_{k\in S}X_i^k(\theta-\delta))(\sum_{k\in S'}X_i^k(\delta_2-\delta_1)),$$
and $$T_3=\frac{1}{n'}\sum_{i=1}^{n'}(\sum_{k\in S'}X_i^k\delta_1)(\sum_{k\in S'}X_i^k(\delta_1-\delta_2)),$$
so that the denominator is $T_1$ and the numerator is $T_2-T_3$. We firstly notice that $\EE[T_1]=\frac{1}{3}s(\delta_1-\delta_2)^2$, and $\EE[T_3]=\frac{1}{3}s(\delta_1-\delta_2)\delta_1$.
Although $\EE[T_2]=0$, but we will show that $T_2$ can be as large as the order
$s\delta_1(\delta_1-\delta_2)$. By Chebyshev's inequality,
\begin{align}
   \Pr\big(|\frac{1}{n'}\sum_{i=1}^{n'}(Y_i-\theta\sum_{k\in S}X_i^k)(\sum_{k\in S'}X_i^k(\delta_2-\delta_1))|>t\big)&\leq \frac{\EE[(Y-\theta\sum_{k\in S}X^k)^2]\EE[(\sum_{k\in S'}X^k)^2](\delta_2-\delta_1)^2}{n't^2}\nonumber\\
   &\leq\frac{s(\delta_2-\delta_1)^2(\Phi^2+\sigma^2)}{n't^2},
\end{align}
so we have $\frac{1}{n'}\sum_{i=1}^{n'}(Y_i-\theta\sum_{k\in S}X_i^k)(\sum_{k\in S'}X_i^k(\delta_2-\delta_1))=O_p(\frac{\sqrt{s(\Phi^2+\sigma^2)}(\delta_1-\delta_2)}{\sqrt{n'}})$. Therefore, we can see that as long as $\sqrt{\frac{\Phi^2+\sigma^2}{n'}}\gg \sqrt{s}\delta_1=\rho\sqrt{\frac{s\log p}{n'}}$, or with another sufficient condition that $\Phi^2\gg s\rho^2\log p$, $|T_2|$ can be as large as $T_3$. Then there is a non-vanishing probability that $T_2+T_3 \leq0 $. In this case, the optimal weight $\hat \beta$ is 0, which indicates $\hat \btheta_L$ is more favorable. However, it is clear to see that the estimation error of $\hat \btheta_L$ is of larger order than that of $\hat \btheta_{SL}$.

At last, we show an example where $\Phi^2\gg s\rho^2\log p$. Set $f(X)=(\sum_{k\in S}X^k)^3$, and denote $A=\frac{\sum_{k\in S}X^k}{\sqrt{s}}$, then we know $\Var(A)=\frac{1}{3}$ and $\sqrt{s}A$ follows a Irwin-Hall Distribution so that $\sqrt{3}A$ can be approximate well by a standard Gaussian variable. Hence,  $\theta=\EE[(\sum_{k\in S}X^k)^4]/\EE[(\sum_{k\in S}X^k)^2]=s\frac{\EE[A^4]}{\EE[A^2]}$ should be of order $s$ and $\Phi^2=\EE[(\sum_{k\in S}X^k)^3-\theta(\sum_{k\in S}X^k)^2]=s^3\EE[(A^3-\frac{\theta}{s} A)^2]$ is of order $s^3$.

{
\subsubsection{Further intuition on modified score function}
Recall that the score function of the supervised lasso/Dantzig selector can be rewritten as
$$
S_1=\frac{1}{n}\sum_{i=1}^n X_i(Y_i-X_i^T\btheta^*)=\frac{1}{n}\sum_{i=1}^n X_i(Y_i-f(X_i))+\frac{1}{n}\sum_{i=1}^nX_i(f(X_i)-X_i^T\btheta^*).
$$
With unlabeled data, we propose to replace the last term with $\frac{1}{n+N}\sum_{i=1}^{n+N}X_i(f(X_i)-X_i^T\btheta^*)$, the sample average over  both labeled and unlabeled data. This leads to the following modified score function
\begin{equation}
S_2=\frac{1}{n}\sum_{i=1}^n X_i(Y_i-f(X_i))+\frac{1}{n+N}\sum_{i=1}^{n+N}X_i(f(X_i)-X_i^T\btheta^*).
\end{equation}	
To see why this manipulation is useful without going through the technical details (e.g, concentration inequalities), let us compare the variance of $S_1$ and $S_2$. Under mild assumptions (e.g. data are i.i.d, $\epsilon$ and $X$ are independent), it is easily seen that
$$
\Cov(S_1)=\frac{1}{n}(\sigma^2+\Phi^2)\bSigma,~\textrm{and}~\Cov(S_2)=\frac{1}{n}\sigma^2\bSigma+\frac{1}{n+N}\Phi^2\bSigma,
$$
where $\sigma^2=\Var(\epsilon)$, $\Phi^2=\EE(f(X)-X^T\btheta^*)^2$ and $\bSigma=\Cov(X)$. Clearly, $\Cov(S_1)-\Cov(S_2)$ is always a semi-positive definite matrix, which implies that the variability of the proposed modified score function is smaller than the original score function. In particular, by comparing the two terms $\frac{1}{n+N}\Phi^2\bSigma$ and $\frac{1}{n}\Phi^2\bSigma$ in $\Cov(S_2)$ and $\Cov(S_1)$ respectively, we can claim that our manipulation can reduce the variance due to model misspecification.
}

{
\subsection{Examples of $\hat h$}\label{app_est_f}

{\bf Sparse additive models}. Assume that $Y_i$ given $X_i=(X_{i1},...,X_{ip})$ follows the additive model
$$
Y_i=\mu+\sum_{j=1}^p f_j(X_{ij})+\epsilon_i,
$$
where $\mu$ is an intercept, $f_j$'s are unknown functions and $\epsilon_i$ is the random error. For identification purpose, we assume $\EE f_j(X_{ij})=0$. Suppose that some of the functions $f_j$ are 0. We also assume $f_j$ is $\bar\ell$-smooth, that is the $k$th derivative $f_j^{(k)}$ exists and satisfies the Lipschitz condition $|f_j^{(k)}(t_1)-f_j^{(k)}(t_2)|\leq C|t_1-t_2|^{\bar\ell-k}$, where $C$ is a constant and $k$ is the greatest integer strictly less than $\bar\ell$.

The literature on how to estimate these additive functions is vast, see \cite{lin2006,meier2009,huang2010,additive} among many others. Here, we adopt the adaptive group lasso approach in \cite{huang2010} as a concrete example. Suppose $X_j$ takes values in $[a,b]$. Let $a=c_0<c_1<....<c_{K+1}=b$ be a partition of $[a,b]$ into $K$ subintervals $I_{t}=[c_t,c_{t+1})$, where $K=n^v$ for some $0<v<1/2$ and $\max_{k}|c_k-c_{k-1}|=O(n^{-v})$. Let $S_n$ denote the space of polynomial splines of order $\ell$, that is the restriction of the function to $I_t$ is a polynomial of degree $\ell$ and for $\ell\geq 2$ the function is $\ell-2$ times continuously differentiable on $[a,b]$. For the space of splines $S_n$, there exists a normalized B-spline basis $\{\phi_k, 1\leq k\leq m_n\}$, where $m_n=K+\ell$, such that for any $f\in S_n$ we have $f(x)=\sum_{k=1}^{m_n}\beta_{k}\phi_k(x)$.

Based on the spline theory, any $\bar\ell$-smooth function $f_j$ can be  approximated by  $\sum_{k=1}^{m_n}\beta_{jk}\phi_k(x)$. Thus, \cite{huang2010} proposed to estimate the unknown functions by solving the following group lasso problem
$$
\min \sum_{i=1}^n \Big[Y_i-\mu-\sum_{j=1}^p\sum_{k=1}^{m_n}\beta_{jk}\phi_k(X_{ij})\Big]^2+\lambda_n\sum_{j=1}^pw_j\|\bbeta_j\|_2,
$$
where $\lambda_n$ is the tuning parameter, $w_j$ is weight in the adaptive lasso, and $\bbeta_j=(\beta_{j1},...,\beta_{jm_n})$. Let us denote the resulting estimator by $\hat\mu$ and $\hat\beta_{jk}$ for $1\leq j\leq p$ and $1\leq k\leq m_n$. Then we have
$$
\hat h(x)=\hat\mu+\sum_{j=1}^p\sum_{k=1}^{m_n}\hat\beta_{jk}\phi_k(x_j).
$$
Under the conditions detailed in \cite{huang2010}, their Corollary 2 implies that $\norm{\hat h-f}_2=O_p(s^{1/2}n^{-\bar\ell/(2\bar\ell+1)})$, where  $\bar\ell$ is the smoothness of the function $f_j$ and $s$ is the number of nonzero $f_j$'s. 

{\bf Pairwise interaction models}. \cite{zhao2016analysis} considered the following interaction models,
$$
Y_i=\sum_{j=1}^p\gamma_{j}X_{ij}+\sum_{1\leq j\leq k\leq p}\gamma_{jk}X_{ij}X_{ik}+\epsilon_i
$$
where $\epsilon_i$ is the random error and $\{\gamma_j\}_{1\leq j\leq p}$ and $\{\gamma_{jk}\}_{1\leq j\leq k\leq p}$ are unknown parameters. Assume that $\{\gamma_j\}_{1\leq j\leq p}$ and $\{\gamma_{jk}\}_{1\leq j\leq k\leq p}$ are sparse. \cite{zhao2016analysis} proposed the following lasso estimator
$$
\min \sum_{i=1}^n \Big[Y_i-\sum_{j=1}^p\gamma_{j}X_{ij}+\sum_{1\leq j\leq k\leq p}\gamma_{jk}X_{ij}X_{ik}\Big]^2+\lambda_1\sum_{j=1}^p|\gamma_j|+\lambda_2\sum_{1\leq j\leq k\leq p}|\gamma_{jk}|.
$$
Let us denote the resulting estimator by $\hat\gamma_j$ and $\hat\gamma_{jk}$. Then we have
$$
\hat h(x)=\sum_{j=1}^p\hat\gamma_{j}x_{j}-\sum_{1\leq j\leq k\leq p}\hat\gamma_{jk}x_{j}x_{k}.
$$
In this case, $\norm{\hat h-f}_2$ corresponds to the prediction error in the lasso problem. Following the proof in \cite{zhao2016analysis}, it can be shown that $\norm{\hat h-f}_2=O_p(\sqrt{s\log p/n})$, where $s$ is the number of nonzero parameters.

We also note that in the analysis of interaction models, one common assumption is known as the strong hierarchy principle, that is if an interaction $X_jX_k$ exists then their main effects (the linear terms of $X_j$ and $X_k$) also exist.  \cite{zhao2016analysis} also proposed a group lasso approach to accommodate the strong hierarchy principle in  interaction models.


}

\subsection{Supplementary Simulation Results}\label{app:plots}

\begin{figure}
    \centering
    \includegraphics[scale=0.55]{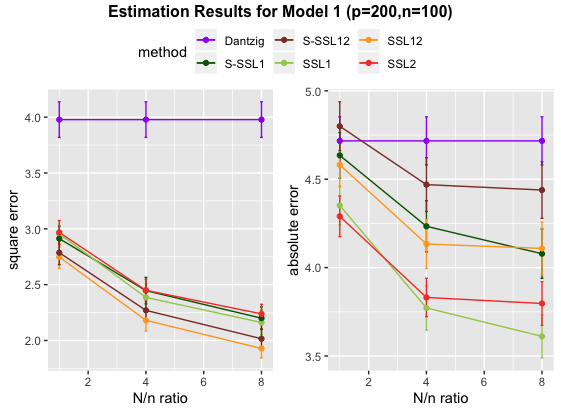}
    \caption{The $L_2$ and $L_1$ estimation error under Model 1 with $p=200$ and $n=100$. The length of the vertical bar represents the magnitude of the sample standard deviations. $L_2$ errors for U-Dantzig are 9.14(0.32), 11.16(0.34), 11.47(0.35). $L_1$ errors for U-Dantzig are 7.68(0.20), 8.33(0.19), 8.49(0.19).
    }
    \label{Fig:EM1p2}
\end{figure}

\begin{figure}
    \centering
    \includegraphics[scale=0.55]{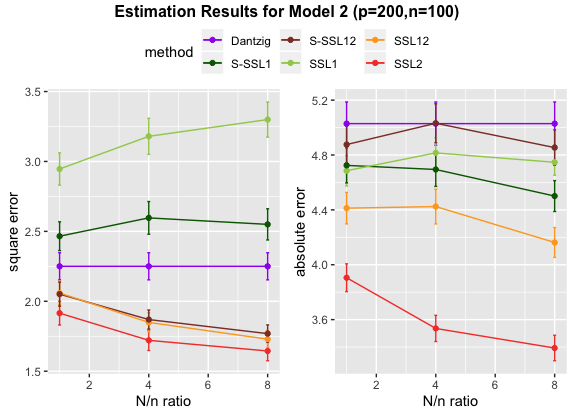}
    \caption{The $L_2$ and $L_1$ estimation error under Model 2 with $p=200$ and $n=100$. The length of the vertical bar represents the magnitude of the sample standard deviations. $L_2$ errors for U-Dantzig are 4.21 (0.14), 4.25(0.14), 4.24(0.14). $L_1$ errors for U-Dantzig are 5.32(0.16), 5.29(0.15), 5.50(0.15).
    }
    \label{Fig:EM2p2}
\end{figure}

\begin{figure}
    \centering
    \includegraphics[scale=0.55]{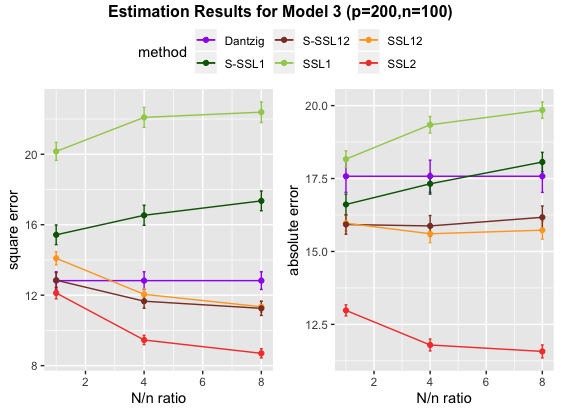}
    \caption{The $L_2$ and $L_1$ estimation error under Model 3 with $p=200$ and $n=100$. The length of the vertical bar represents the magnitude of the sample standard deviations.
    $L_2$ errors for U-Dantzig are 17.46(0.45), 19.03(0.40), 19.73(0.44). $L_1$ errors for U-Dantzig are 16.99(0.27), 17.70(0.26), 17.88(0.31).
    }
    \label{Fig:EM3p2}
\end{figure}

\begin{figure}
    \centering
    \includegraphics[scale=0.55]{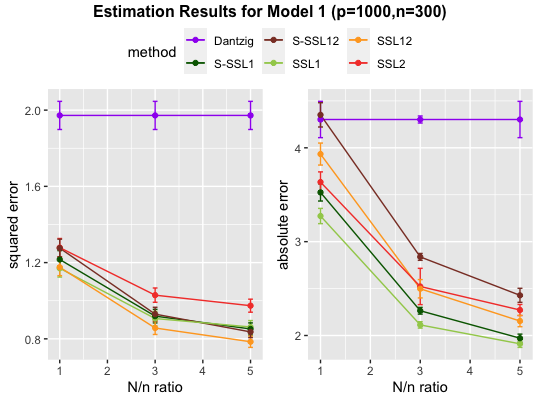}
    \caption{The $L_2$ and $L_1$ estimation error under Model 1 with $p=1000$ and $n=300$. The length of the vertical bar represents the magnitude of the sample standard deviations. $L_2$ errors for U-Dantzig are 4.12(0.12), 4.94(0.14), 5.18(0.15). $L_1$ errors for U-Dantzig are 5.09(0.11), 5.57(0.10), 5.74(0.12).
    }
    \label{Fig:EM1p10}
\end{figure}

\begin{figure}
    \centering
    \includegraphics[scale=0.55]{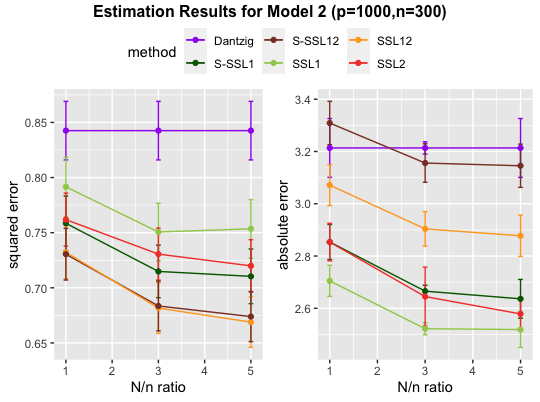}
    \caption{The $L_2$ and $L_1$ estimation error under Model 2 with $p=1000$ and $n=300$. The length of the vertical bar represents the magnitude of the sample standard deviations. $L_2$ errors for U-Dantzig are 1.82(0.06), 2.01(0.06), 2.11(0.07). $L_1$ errors for U-Dantzig are 3.47(0.06), 3.89(0.07), 3.98(0.07).
    }
    \label{Fig:EM2p10}
\end{figure}

\begin{figure}
    \centering
    \includegraphics[scale=0.55]{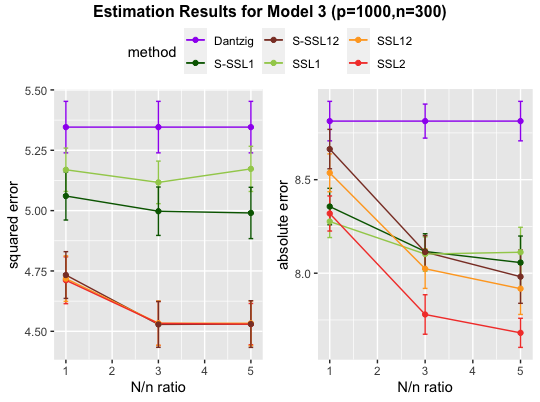}
    \caption{The $L_2$ and $L_1$ estimation error under Model 3 with $p=1000$ and $n=300$. The length of the vertical bar represents the magnitude of the sample standard deviations.
    $L_2$ errors for U-Dantzig are 8.82(0.18), 10.32(0.21), 10.77(0.21). $L_1$ errors for U-Dantzig are 11.81(0.14), 13.97(0.19), 14.33(0.20).
    }
    \label{Fig:EM3p10}
\end{figure}

{
Parallel to the results presented in the main paper for Models 1, 2, 3 with $p=500$ and $n=200$, here we also present the similar results for Models 1, 2, 3 with $p=200$ and $n=100$ in Figures~\ref{Fig:EM1p2}, \ref{Fig:EM2p2} and \ref{Fig:EM3p2}, as well as for Models 1, 2, 3 with $p=1000$ and $n=300$ in Figures~\ref{Fig:EM1p10}, \ref{Fig:EM2p10} and \ref{Fig:EM3p10}, respectively.
The results from analyzing the $L_2$ and $L_1$ estimation errors are similar to the main paper, so are omitted.

When creating the aggregated estimator $\hat \btheta_{AH}$ in Section~\ref{sec_safe2}, suppose that we have
two candidate models $h_1$ and $h_2$ for the conditional mean functions. The aggregated estimator in (\ref{est_ag}) is given by $\hat\btheta_{AH}=\hat\btheta_{h_1}+\hat\bomega$, where we use the semi-supervised estimator with model $h_1$, $\hat\btheta_{h_1}$, as an initial estimator.
The creation of $\hat \btheta_{AH}$ is not symmetric to $h_1$ and $h_2$.
Theoretically, Proposition 5.1 shows that the aggregated estimators have the same convergence rate regardless whether we use $\hat\btheta_{h_1}$ or $\hat\btheta_{h_2}$ as the initial estimator.
To back up our theory, we add new simulation results for the aggregated estimator named as SSL21, which uses $\hat\btheta_{h_2}$ as the initial estimator.
Tables \ref{M1_order},\ref{M2_order},\ref{M3_order} show that the two aggregated estimators SSL21 and SSL12 have similar performance.

Additionally, in Section~\ref{sec:computebxi}, we adopt the splitting strategy 50-50 in the cross-fitting technique; i.e., we use 50\% of labeled data to train the model $\hat h^{-j}$ and the other 50\% to compute $\hat \bxi_j$; see (\ref{eq_hatxi}).
It is of interest to consider an imbalanced split, such as 70-30.
Note that, due to the symmetry of calculating the estimator $\hat\btheta_{SD}$ in (\ref{est}), the split strategies of 70-30 and 30-70 essentially result in the same estimator.
The comparison results in Table~\ref{table_cf} show that the splitting strategy of 50-50 is slightly preferred.

}

\begin{table}
\centering
\begin{tabular}{|c|c|c|c|c|c|c|}
\hline
 &\multicolumn{2}{c|}{N=300 }&\multicolumn{2}{c|}{N=900} & \multicolumn{2}{c|}{N=1500}\\
\cline{2-7}
&$L_1$ error &$L_2$ error& $L_1$ error &$L_2$ error &$L_1$ error &$L_2$ error\\
\cline{1-7}
SSL12 &3.93& 1.17& 2.50& 0.86& 2.15 & 0.78 \\
\hline
SSL21 &3.84  & 1.19 & 2.40 &0.86& 2.10 & 0.79\\
\hline
\end{tabular}
\caption{The $L_1$ and $L_2$ estimation errors of SSL12, SSL21 with Model 1 for $n=300$, $p=1000$.}
\label{M1_order}
\end{table}

\begin{table}
\centering
\begin{tabular}{|c|c|c|c|c|c|c|}
\hline
 &\multicolumn{2}{c|}{N=300 }&\multicolumn{2}{c|}{N=900} & \multicolumn{2}{c|}{N=1500}\\
\cline{2-7}
&$L_1$ error &$L_2$ error& $L_1$ error &$L_2$ error &$L_1$ error &$L_2$ error\\
\cline{1-7}
SSL12 &3.07& 0.73& 2.90& 0.68& 2.88 & 0.67 \\
\hline
SSL21 &2.98  & 0.73 & 2.70 &0.67& 2.63 & 0.66\\
\hline
\end{tabular}
\caption{The $L_1$ and $L_2$ estimation errors of SSL12, SSL21 with Model 2 for $n=300$, $p=1000$.}
\label{M2_order}
\end{table}

\begin{table}
\centering
\begin{tabular}{|c|c|c|c|c|c|c|}
\hline
 &\multicolumn{2}{c|}{N=300 }&\multicolumn{2}{c|}{N=900} & \multicolumn{2}{c|}{N=1500}\\
\cline{2-7}
&$L_1$ error &$L_2$ error& $L_1$ error &$L_2$ error &$L_1$ error &$L_2$ error\\
\cline{1-7}
SSL12 &8.54& 4.72& 8.02& 4.53& 7.92 & 4.53 \\
\hline
SSL21 &8.39  & 4.71 & 7.81&4.51& 7.70 & 4.51\\
\hline
\end{tabular}
\caption{The $L_1$ and $L_2$ estimation errors of SSL12, SSL21 with Model 3 for $n=300$, $p=1000$.}
\label{M3_order}
\end{table}

\begin{table}
\centering
\begin{tabular}{|c|c|c|c|c|c|c|c|}
\hline
& &\multicolumn{2}{c|}{Model 1} &\multicolumn{2}{c|}{Model 2} & \multicolumn{2}{c|}{Model 3}\\
\cline{2-8}
& split ratio &SSL1& SSL2 &SSL1& SSL2 &SSL1 &SSL2 \\
\cline{1-8}
\multirow{2}{*}{$L_2$ error}&50-50 &0.74 &1.03 &0.80 &0.65 &5.65 &4.25\\
     \cline{2-8}
    &70-30 & 0.78&1.08 &0.96 &0.74 &6.06 &4.55 \\
\hline
\multirow{2}{*}{$L_1$ error}&50-50 &2.20 &2.53 &2.97&2.62 &9.06 &7.77\\
   \cline{2-8}
    &70-30 &2.44 & 2.74&3.55 &2.62 &9.30 &8.07 \\
\hline
\end{tabular}
\caption{The $L_1$ and $L_2$ estimation errors of estimators SSL1 and SSL2 with split ratio 50-50 and 70-30, for the setting with $p=1000$, $n=400$ and $N=1200$.}
\label{table_cf}
\end{table}

\subsection{Supplementary Materials in Real Data Application}\label{app:plotsdata}

{

The MIMIC-III database is comprehensive in nature and it includes 26 individual tables in csv format, named “admissions”, “patients”, “outputevents”, “chartevents”, etc. It includes vital signs, medications, laboratory measurements, observations and notes charted by care providers. The total data size reached 40GB after decompression. There are two types of data in the database: static data and dynamic data. The static data do not change over time and are recorded only once such as the date of birth, while the dynamic data are mainly the vital signs that periodically measured during the patient’s ICU stay such as the blood pressure and the heart rate.
The Table 3 of \cite{johnson2016mimic} contains all the classes of the data available in MIMIC-III while their Table 4 provides an overview of the 26 data tables.

Due to the complexity of this large database, a few preprocessing steps are conducted to improve the quality of the data.

For the patient selection process, we exclude patients with less than 24-hour ICU stays, since those patients with short length of stays have many missing values in the covariates. Patients who died during the ICU stay are also excluded.
Additionally, for patients with multiple ICU admissions, we only keep the first ICU admission in our analysis.
This database has a variety of missing values with different patterns, including for the outcome ``albumin''.
To illustrate our method, we choose not to impute this outcome and instead only focus on the patients whose ``albumin'' is available.

We consult the previous literature \citep{brown2012epidemiology, tabak2017predicting, xue2019predicting, liu2020serum, du2021prediction, jhou2021plasma, tang2021association} to select the variables used in our data analysis.
In general, the variable selection takes into account the following directives \citep{fialho2012data, nates2016icu}.
First, variables must be easily and/or routinely assessed in the 24 hours before discharge.
There is a special concern for the patients who were older than 89 years, because their dates of birth were shifted to obscure their true age to comply with Health Insurance Protability and Accountability Act (HIPAA) regulations. These patients appear in the database with ages of over 300 years, and the MIMIC-III team only provides the median age for the patients before shifting, which is 91.4.
Thus, for the patients with shifted age, we replace their age with the median 91.4.
Second, the balance should be kept in the number of selected variables given that it will affect the number of patients that will be used in the analysis, i.e., the more variables that are defined, the fewer patients are likely to have all of them collected at the same time; or the high number of variables will bias the dataset towards selecting patients having similar conditions.
Third, the variables selected should be independent with minimum correlation.
For dynamic data such as many clinical biomarkers with continuous scale, the database collects the minimum, the maximum, as well as the mean, values across a certain period of time.
In order to alleviate the potential collinearity among these variables, we decide to only include the mean values in our analysis.
Additionally, besides the 78 covariates themselves, we include pairwise interactions among 75 of the covariates that are continuous-scale clinical biomarkers as well as their squared terms, which all together result totally $p=2928$ variables.

Regarding other missing data issues, in our situation,
around 54\% covariates contain missing values. Among these covariates with missing values, the missingness proportions are 9.4\% on average and the range is from 0.2\% to 30.8\%.
For the demographic variables, most observations were complete.
For the laboratory test observations, observations are missing mainly because the variables are considered to be irrelevant to the current clinical problems, thus those laboratory tests are not requested by clinician.
For those missing values, we simply impute them using the mean of observed samples, the so-called mean imputation.

After all of these data pre-processing steps, we are left with a dataset with 4784 patients and each of them has $p=2928$ covariates.
In our analysis, we consider a high dimensional setting where the sample size of our labeled data is $n=500$.

Last but not least, regarding our analysis results, we also collect the variable selection results from the comparison of supervised lasso, S-SSL and SSL estimators.
The results show that the variables \verb"white blood cells", \verb"hemoglobin" and \verb"lymphocytes" are selected more frequently by both of SSL and S-SSL than supervised lasso. In the medical literature, there are various evidence documenting the associations between the albumin level in the blood sample with our selected biomarkers here, such as with the hemoglobin level \citep{fukui2008association}, with the lymphocytes level \citep{alagappan2018albumin}, and with the count of the white blood cells \citep{cavalot2002white}.
More interestingly, researchers recently started to use the so-called HALP score, a combination of hemoglobin, albumin, lymphocyte, and platelet levels in the blood sample \citep{chen2015prognostic}, as a prognostic factor for various disease types \citep{peng2018prognostic, guo2019hemoglobin, shen2019hemoglobin, tojek2019blood}. By  exploring the data with the random forest model, our statistical results from the semi-supervised methods are consistent with the medical literature.

}

\subsection{Discussion}\label{sec_disc}

In this paper, we consider high-dimensional semi-supervised learning, and focus on when and how we can exploit the unlabeled data to achieve the optimal and safe estimates of model parameters. Our key observation is that, the covariate $X$ in the unlabeled data, whose sample size can be much bigger then the labeled data, could be very informative for the parameter of interest in a misspecified model.
We derive the minimax lower bound for parameter estimation in the semi-supervised setting, and show that generally the traditional supervised estimators without using the unlabeled data cannot attain this lower bound.
Then, we propose a semi-supervised estimator, which depends on the correct specification of the conditional mean function, can attain the minimax lower bound and hence is optimal.
To alleviate the strong requirement for this optimal estimator, we further propose a safe semi-supervised estimator. We view it safe, because this estimator remains minimax optimal when the conditional mean function is correctly specified, and is always at least as good as the supervised estimators.

While our focus is on the parameter estimation, we should note some recent and important progress in inference, particularly in the construction of confidence intervals with high-dimensional data in the semi-supervised setting\citep{2019jelena,caiexplained}. It would be important and challenging to develop procedures for statistical inference under our framework.

\bibliographystyle{asa}
\bibliography{reference}